\documentclass[twocolumn,showpacs,preprintnumbers,amsmath,amssymb]{revtex4}

\usepackage{graphicx}
\usepackage{dcolumn}
\usepackage{bm}

\begin{document}


\title{The Magnetic Force as a Consequence of the Thomas Precession}


\author{David C. Lush }
\affiliation{%
d.lush@comcast.net \\}%


\date{\today}

\begin{abstract}

The requirements imposed by relativistic covariance on the physical description of two interacting classical charged particles are investigated. Because rotational pseudoforces cannot be caused by Thomas precession,  kinematical considerations demand the presence of compensatory forces when Thomas precession of an inertial reference frame is observed.  The magnetic force on a moving charge is apparently one such force, where Thomas precession of the laboratory frame is seen by an observer co-moving with the charge.  Thus, no acceleration of the field source charge is required to cause the necessary Thomas precession, consistent with the known properties of the magnetic interaction. However, when the field source charge is accelerating, an additional magnetic-like force is expected. Other forces corresponding to the Euler and centrifugal rotational pseudoforces are also predicted by this line of reasoning. The plausibility that an anti-centrifugal force of the Thomas precession may account for the binding of quarks into nucleons is investigated. The similarity of the magnetic force on a relativistically-moving charge in the radiative magnetic field of a nearby Coulomb-accelerating charge to the predicted anticentrifugal force of the Thomas precession is shown. 
\end{abstract}

\pacs{03.30.+p, 41.20.-q, 45.05.+x, 45.20.da}


\maketitle

\section{Introduction}

The rest frames of charged particles interacting electrodynamically are known to rotate relative to the laboratory inertial reference frame, as well as mutually, due to the Thomas precession \cite{Thomas1927}. Kinematics requires that when a reference frame rotates relative to an inertial reference frame, rotational pseudoforces must be present in the rotating frame.  However, it is clear that Thomas precession cannot cause rotational pseudoforces in general, because it depends only on the relative motion between two reference frames, and the Thomas precessing frame may be an inertial frame if the observer is accelerating.  Thus, for example, since there can be no Coriolis force in the rest frame of a field-source charged particle that is only translating with constant velocity, an observer co-moving with a charged test particle that is translating relative to the field-source particle, and cross-accelerating relative to the translation, must infer the presence of forces that compensate for the lack of Coriolis force the observer expects due to the observed Thomas precession.  

It is contended in the present contribution that the necessity of an observer moving non-inertially to infer pseudoforce-compensating forces implies even in inertial frames the existence of related forces, that will be herein referred to as anti-pseudoforces, the most obvious of which can be recognized as the magnetic part of the Lorentz force. The formal similarity of the magnetic force to a Coriolis force is viewed as a direct consequence of it arising as an anti-Coriolis force.  Similar reasoning when extended to centrifugal and Euler pseudoforces implies the existence of other tangible forces.  These may provide relativistic kinematical bases for the strong and weak forces. Also, when a magnetic field source charge is accelerating as well as translating, an additional magnetic-like force must be expected. It is shown further that under highly relativistic and short range conditions, the force on a charged particle due to the magnetic acceleration field can behave similarly to predicted anticentrifugal force.  That is, the magnetic field of an accelerating charge can cause a force between two charges that is attractive independent of the relative polarity of the charges, and can overcome Coulombic repulsion at sub-nucleonic scale in the highly relativistic limit.

\section{Similarity of the Magnetic Force to a Coriolis Force}

In this section the approximate Lorentz force on a charged test particle interacting electromagnetically with another charged particle is determined and the magnetic part of the interaction identified, and then it is shown how the magnetic force may be interpreted as an anti-Coriolis force.  The anti-Coriolis force acts in any inertial reference frame where both the field source particle and test particle are moving.

\subsection{Interaction of Two Charged Particles, Where One Particle is Non-Accelerating}

The relativistic law of inertia for a massive particle of momentum \(\mbox{\boldmath$P$}\) and rest mass \(m\), acted on by a force \(\mbox{\boldmath$F$}\) is
\begin{equation}
\mbox{\boldmath$F$} = \frac{d\mbox{\boldmath$P$}}{dt} =\frac{d}{dt} \left[
\gamma m \mbox{\boldmath$v$}\right] = \dot{\gamma} m \mbox{\boldmath$v$} + \gamma m \mbox{\boldmath$a$},
\label{RLOM}
\end{equation}
where \(\mbox{\boldmath$v$}\) is the particle velocity and \(\mbox{\boldmath$a$} \equiv d\mbox{\boldmath$v$}/dt \equiv \dot{\mbox{\boldmath$v$}}\) its acceleration, and \(\gamma \equiv 1/\sqrt{(1-(v/c)^2}\) with \(v \equiv |\mbox{\boldmath$v$}|\) and c the speed of light. The acceleration due to \(\mbox{\boldmath$F$}\) is thus
\begin{equation}
\mbox{\boldmath$a$} = \left[\frac{1}{\gamma m} \right]\left[\mbox{\boldmath$F$} - \dot{\gamma} m \mbox{\boldmath$v$} \right] = \left[\frac{1}{\gamma m} \right]\left[\mbox{\boldmath$F$} - \gamma^3(\mbox{\boldmath$\beta$} \cdot\dot{\mbox{\boldmath$\beta$}}) m \mbox{\boldmath$v$} \right],
\label{accel_from_RLOM}
\end{equation}
where \(\mbox{\boldmath$\beta$} \equiv \mbox{\boldmath$v$}/c\) (and using the well-known identity that \(\dot{\gamma} = \gamma^3(\mbox{\boldmath$\beta$} \cdot\dot{\mbox{\boldmath$\beta$}})\)). The force of interest is the Lorentz force on a moving charged particle in the electromagnetic field caused by another moving charged particle.  In this case the electromagnetic field is exactly described by the Li\'enard-Wiechert field expressions \cite{jcksn:classelec}.  The Li\'enard-Wiechert field expressions in three-vector notation are 
\begin{widetext}
\begin{equation}
\mbox{\boldmath$E$}(\mbox{\boldmath$r$},t) = q \left[ \frac{\mbox{\boldmath$n$}-\mbox{\boldmath$\beta$}}{\gamma^2 \left( 1 - \mbox{\boldmath$\beta$} \cdot \mbox{\boldmath$n$}\right)^3 R^2 }\right]_{\text{ret}} + \frac{q}{c} \left[ \frac{\mbox{\boldmath$n$} \times \left((\mbox{\boldmath$n$}-\mbox{\boldmath$\beta$})\times \dot{\mbox{\boldmath$\beta$}}\right)}{\left(1 - \mbox{\boldmath$\beta$} \cdot \mbox{\boldmath$n$}\right)^3 R }\right]_{\text{ret}}
\label{LW_E_field}
\end{equation}
\end{widetext}
and
\begin{equation}
\mbox{\boldmath$B$}(\mbox{\boldmath$r$},t) = \left[ \mbox{\boldmath$n$} \times \mbox{\boldmath$E$}  \right]_{\text{ret}}, 
\label{LW_B_field}
\end{equation}
where, if \(\mbox{\boldmath$r$}\) is the displacement from a field-source charged particle at the retarded time \(t' \equiv t - R/c\) to a field point at time \(t\), then \(R \equiv |\mbox{\boldmath$r$} |\),  and \(\mbox{\boldmath$n$} = \mbox{\boldmath$r$}/R\).  Also \(\mbox{\boldmath$\beta$}  \equiv \mbox{\boldmath$v$}/c\), where \(\mbox{\boldmath$v$}\) is the field-source particle velocity. The subscript ``ret'' refers to that the quantity in the brackets is evaluated at the retarded time.  

The first term on the right hand side of Eq. (\ref{LW_E_field}) is called the velocity, or non-radiative, electric field.  The second is called the acceleration or radiative electric field.  In the present application, with the electromagnetic field caused by a non-accelerating charge, the acceleration fields vanish identically. It can also be seen by inspection that the magnitude of the field difference from the Coulomb field of the particle (that is, the electric field in the rest frame of the particle) due to motion is small when \(\beta << 1\).  Since the magnetic force strength is generally of order \((v/c)^2\) compared to the Coulomb force, it will be sufficient to represent all forces and fields only to this order.  

From (\ref{LW_E_field}) and with the source particle non-accelerating, the electric field is exactly
\begin{equation}
\mbox{\boldmath$E$}(\mbox{\boldmath$r$},t) = q_s \left[ \frac{\mbox{\boldmath$n$}-\mbox{\boldmath$\beta$}_s}{{\gamma_s}^2 \left( 1 - \mbox{\boldmath$\beta$}_s \cdot \mbox{\boldmath$n$}\right)^3 R^2 }\right]_{\text{ret}} 
\label{LW_E_vel_field}
\end{equation}
and the magnetic field is approximated to order \((v/c)^2\) as
\begin{equation}
\mbox{\boldmath$B$}(\mbox{\boldmath$r$},t) \approx \frac{q_s}{R^2} \left[\mbox{\boldmath$\beta$}_s \times\mbox{\boldmath$n$}\right].  
\label{B_field}
\end{equation}

Let \(\mbox{\boldmath$r$}_s\) and  \(\mbox{\boldmath$r$}_t \) represent position vectors to an electromagnetic field source particle of mass \(m_s\), and a test particle of mass \(m_t\), in an inertial reference frame (IRF) that will be referred to herein as the laboratory frame, and \(R\equiv|\mbox{\boldmath$r$}|\equiv|\mbox{\boldmath
$r$}_{t} - \mbox{\boldmath $r$}_{s}|\).  The velocities of both the source and test particles are assumed nonvanishing in the laboratory frame. It is also assumed for simplicity that the particles have no intrinsic magnetic moments.  The Lorentz force on the test particle in the laboratory frame is then
\begin{equation}
\mbox{\boldmath$F$} = \frac{q_t\mbox{\boldmath$v$}_t}{c} \times  \mbox{\boldmath$B$} + q_t \mbox{\boldmath$E$}. 
\label{TotalLorentzForce}
\end{equation}

For the uniformly translating field source particle, it is straightforward to evaluate the retardation effects explicitly and rewrite Eq. (\ref{TotalLorentzForce}) in terms of non-retarded quantites.  The result accurate to order \((v/c)^2\) is
\begin{equation}
\mbox{\boldmath$F$} \approx \frac{q_s q_t}{R^2}\left[   \mbox{\boldmath$\beta$}_t \times\left[\mbox{\boldmath$\beta$}_s \times\mbox{\boldmath$n$}\right]\right]  +  \frac{q_s q_t\mbox{\boldmath$n$}}{\left(1 - 3(\mbox{\boldmath$\beta$}_s \cdot \mbox{\boldmath$n$})^2/2 + 3{\beta_s}^2/2 \right) R^2}.
\label{Ftotal}
\end{equation}

If the force is considered as consisting of electric and magnetic parts so that \(\mbox{\boldmath$F$} \equiv \mbox{\boldmath$F$}_{\text{electric}} + \mbox{\boldmath$F$}_{\text{magnetic}}\), it will be useful to further consider the electric part of the force as consisting of a Coulomb force plus additional terms that are at order \(\beta\) and higher powers of \(\beta\), where the Coulomb part of the electric force is
\begin{equation}
\mbox{\boldmath$F$}_{\text{Coul}} \equiv \frac{q_t q_s}{ R^3}(\mbox{\boldmath$r$}_{t} - \mbox{\boldmath$r$}_{s}) = \frac{q_t q_s \mbox{\boldmath$r$}}{R^3}.
\label{F_coul}
\end{equation}

The acceleration of the test particle due to the Coulomb force, using Eq. (\ref{accel_from_RLOM}), is
\begin{equation}
\mbox{\boldmath$a$}_{\text{Coul}} = \left[\frac{1}{\gamma_t m_t} \right]\left[\mbox{\boldmath$F$}_{\text{Coul}} - {\gamma_t}^3(\mbox{\boldmath$\beta$}_t \cdot\dot{\mbox{\boldmath$\beta$}}_t) m_t \mbox{\boldmath$v$}_t \right],
\label{Coul_accel_from_RLOM}
\end{equation}
which to order \((v/c)^2\) is
\begin{equation}
\mbox{\boldmath$a$}_{\text{Coul}} \approx \frac{\mbox{\boldmath$F$}_{\text{Coul}}}{\gamma_t m_t} - (\mbox{\boldmath$\beta$}_t \cdot\dot{\mbox{\boldmath$\beta$}}_t) \mbox{\boldmath$v$}_t,
\label{Coul_accel_from_RLOM_approx}
\end{equation}

The acceleration of the test particle due to the Coulomb force, neglecting the relativistic terms from Eq. (\ref{Coul_accel_from_RLOM_approx}) that are of order \(\beta^2\) and higher, is then
\begin{equation}
\mbox{\boldmath$a$}_{\text{Coul}} \approx \frac{q_t q_s \mbox{\boldmath$r$}}{m_t R^3}.
\label{Coulomb_accel}
\end{equation}

If we let \(\mbox{\boldmath$F$} \equiv \mbox{\boldmath$F$}_{\text{electric}} + \mbox{\boldmath$F$}_{\text{magnetic}}\) with
\begin{eqnarray}
\mbox{\boldmath$F$}_{\text{magnetic}} =  q_t \mbox{\boldmath$\beta$}_t \times \mbox{\boldmath$B$}, 
\label{BiotSavartForce}
\end{eqnarray}
then the magnetic force can be written using Eq. (\ref{B_field}) as
\begin{eqnarray}
\mbox{\boldmath$F$}_{\text{magnetic}} \approx  \left[\frac{q_t q_s}{R^2}   \right]\left[ \mbox{\boldmath$\beta$}_t \times (
\mbox{\boldmath$\beta$}_s  \times \mbox{\boldmath$n$} ) \right] = - 2 m_t \mbox{\boldmath$\omega$} \times \mbox{\boldmath$v$}_t,
\label{CoriolisLikeForce}
\end{eqnarray}
with 
\begin{eqnarray}
\mbox{\boldmath$\omega$} =  \left[\frac{q_t q_s}{2 c m_t R^2}   \right]\left[ 
\mbox{\boldmath$\beta$}_s  \times \mbox{\boldmath$n$} \right] \approx  \left[\frac{1}{2 c^2}   \right]\left[ 
\mbox{\boldmath$v$}_s  \times \mbox{\boldmath$a$}_{\text{Coul}} \right]
\label{DerivedAngularVelocity}
\end{eqnarray}
where \(\mbox{\boldmath$a$}_{\text{Coul}} \) is the Coulomb acceleration of the test particle as given by Eq. (\ref{Coulomb_accel}).

Now, Eq. (\ref{CoriolisLikeForce}) shows that the magnetic force on the test particle is formally identical to a Coriolis force that would be present if the reference frame of the description was rotating with angular velocity as given by Eq. (\ref{DerivedAngularVelocity}). Furthermore, the above angular velocity expression can be related to the expression for the Thomas precession  \cite{jcksn:classelec} in the limit of small \(v/c\), of the rest frame of a particle with velocity \( \mbox{\boldmath$v$}\) and acceleration \( \mbox{\boldmath$a$}\) relative to the observer seeing the Thomas precession. That is,
\begin{equation}
\mbox{\boldmath$\omega$}_{\text{T}} = \frac{\gamma^2}{\gamma + 1}\frac{\mbox{\boldmath$a$} \times
\mbox{\boldmath$v$}} {c^2} \approx  \frac{1}{2}\frac{\mbox{\boldmath$a$} \times
\mbox{\boldmath$v$}} {c^2}.  
\label{Thomas_av_approx}
\end{equation}

The acceleration of coordinate axes that are fixed in the laboratory frame, relative to the observer co-moving with the test particle, is simply the opposite of the test particle acceleration, since the laboratory frame is non-accelerating.  Similarly, the relative velocity to the same observer of an object fixed in the laboratory frame is the opposite of the test particle velocity as observed from the laboratory frame.  However, the sign of the Thomas precession must be inverted to account for the observer being in the non-inertial frame.  This results in an angular velocity of Thomas precession of the laboratory frame coordinate axes seen by the test particle co-moving observer of   
\begin{equation}
\mbox{\boldmath$\omega$}'_{\text{T}} \equiv - \mbox{\boldmath$\omega$}_{\text{T}} \approx  -\frac{1}{2}\frac{\mbox{\boldmath$a$}_t \times
\mbox{\boldmath$v$}_t} {c^2},
\label{Thomas_av_decomp}
\end{equation}
which is not equal to the expected angular velocity according to Eq. (\ref{DerivedAngularVelocity}). In order to derive the magnetic force from Thomas precession seen from the test particle, it will be necessary to consider not just the Thomas precession of the laboratory frame, but also that of the field source particle rest frame, in both cases as seen by the test particle co-moving observer.

\subsection{The Magnetic Force as a Coriolis Effect of the Thomas Precession}

In accordance with Eq. (\ref{Thomas_av_approx}), but with a sign inversion to account for the difference in observer, the test particle co-moving observer sees any Cartesian coordinate axes that are fixed in the laboratory frame as Thomas precessing with angular velocity
\begin{equation}
\mbox{\boldmath$\omega$}_l \approx -\frac{1}{2}\frac{\mbox{\boldmath$a$}_t \times
\mbox{\boldmath$v$}_t} {c^2}.
\end{equation}

Similarly, the angular velocity of the Thomas precession of the source particle rest frame seen by the test particle co-moving observer is 
\begin{equation}
\mbox{\boldmath$\omega$}_s \approx -\frac{1}{2}\frac{\mbox{\boldmath$a$}_t \times
(\mbox{\boldmath$v$}_t - \mbox{\boldmath$v$}_s))} {c^2}.
\end{equation}

The relative angular velocity of the laboratory frame compared to the field-source particle rest frame is then
\begin{equation}
\mbox{\boldmath$\omega$}_r =  \mbox{\boldmath$\omega$}_l - \mbox{\boldmath$\omega$}_s \approx  -\frac{1}{2}\frac{\mbox{\boldmath$a$}_t \times
\mbox{\boldmath$v$}_s} {c^2}. 
\label{Thomas_rel_omega}
\end{equation}

The observer co-moving with the test particle thus sees the laboratory frame as rotating with angular velocity \(\mbox{\boldmath$\omega$}_r\) with respect to the field source particle rest frame. Although the test particle co-moving observer sees both frames as rotating, it is the relative rotation that determines the kinematical relationship between them. If the law of motion is known in either of the two frames, it can be determined in the other using standard kinematics.  If the source rest frame is taken as the non-rotating frame, then the test particle co-moving observer predicts that the lab frame equation of motion must be the source frame equation plus the Coriolis, Euler, and centrifugal rotational pseudoforces. These are viewed as anti-pseudoforces since the source frame is rotating with a larger-magnitude angular velocity than the lab frame, from the point of view of the test particle rest frame observer, yet the electromagnetic interaction in the source particle rest frame is perfectly radial in general and so apparently lacking in any rotational pseudoforces.  (This is only to be expected, since an observer co-moving with the uniformly-translating field source particle experiences no Thomas precession.)  

The expected Coriolis force in the lab frame relative to the source particle rest frame, as observed from the test particle rest frame is
\begin{eqnarray}
\mbox{\boldmath$F$}_{\text{Coriolis}} = -2 m_t \mbox{\boldmath$\omega$}_r \times \mbox{\boldmath$v$}_t, 
\label{AntiCoriolisForce}
\end{eqnarray}
or, with the relative angular velocity of the Thomas precession between the source and laboratory frames given by Eq. (\ref{Thomas_rel_omega}),
\begin{eqnarray}
\mbox{\boldmath$F$}_{\text{Coriolis}} =  \frac{m_t}{c^2}\left[\mbox{\boldmath$a$}_t \times
\mbox{\boldmath$v$}_s\right] \times \mbox{\boldmath$v$}_t .
\end{eqnarray}

Approximating the test particle acceleration as that due to the Coulomb force due to the source particle according to Eq. (\ref{Coulomb_accel}) obtains
\begin{eqnarray}
\mbox{\boldmath$F$}_{\text{Coriolis}} =  \frac{q_t q_s}{R^2}  \left[\mbox{\boldmath$\beta$}_t \times \left[\mbox{\boldmath$\beta$}_s \times
\mbox{\boldmath$n$}\right]\right]. 
\label{AntiCoriolisForceRef}
\end{eqnarray}

Comparing with Eq. (\ref{CoriolisLikeForce}), it is apparent that 
\begin{eqnarray}
\mbox{\boldmath$F$}_{\text{Coriolis}} = \mbox{\boldmath$F$}_{\text{magnetic}}.
\end{eqnarray}

The magnetic force on the test particle is thus interpretable as a Coriolis force caused by Thomas precession.

It seems worth remarking that although the Coriolis and other inertial pseudoforces generally are directly proportional to the mass of the object on which they appear to act, the magnetic component of the Coriolis force manifesting here does not depend on the test particle mass.  The amount of Thomas precession seen by the test particle rest frame observer is inversely proportional to the test particle mass, which has directly canceled the mass factor that would be otherwise present.  Such cancelation of the mass factor is of course essential to admitting the possibilty that the magnetic force is due to a Coriolis effect.  However, the argument presented here seems to also be applicable to other situations where such cancelation cannot occur.  Specifically, had the field-source particle been allowed to freely accelerate in the Coulomb field of the test particle, then this acceleration would have contributed to the Thomas precession of the source particle rest frame seen by the observer in the test particle rest frame, with amount of additional Thomas precession depending inversely on the mass of field-source particle.  Thus, an additional Coriolis force component would be present that would be proportional to the ratio of the test particle to source particle masses.  In the case of interactions between equal-mass free particles, this would seem to result in a doubling of the strength of the magnetic interaction.   	     


The magnetic force has been related by other authors to a Coriolis force \cite{Bergstrom1973}, as well as to an anti-Coriolis force \cite{RoyerY11}.  However, those authors do not infer the existence of anti-Euler or anti-centrifugal forces. The apparent incompleteness of the Lorentz force has also been noted previously \cite{Hadjes:Y10}.

\section{Strength of the Anti-Centrifugal Force Compared to Coulomb Repulsion}

The same arguments that lead to expectation of an anti-Coriolis force lead also to expectation of an anti-centrifugal force, that can be given notionally as 
\begin{eqnarray}
\mbox{\boldmath $F$}_{\text{anticentrifugal}} \equiv \mbox{\boldmath $F$}_{\text{a.c.}} \equiv  \gamma m \mbox{\boldmath $\omega$}_{\text{T}} \times \left( \mbox{\boldmath $\omega$}_{\text{T}} \times \mbox{\boldmath $r$} \right),
\label{AntiCentForceDef}
\end{eqnarray}
with (using again the formula for the angular velocity of the Thomas precession given in \cite{jcksn:classelec}, but here without specialization to small \(v/c\)),
\begin{equation}
\mbox{\boldmath$\omega$}_{\text{T}} = - \left(\gamma - 1\right) \frac{\left[\mbox{\boldmath$v$} \times
\mbox{\boldmath$a$}\right]} {{v}^2} \approx - \gamma  \frac{\left[\mbox{\boldmath$v$} \times
\mbox{\boldmath$a$}\right]} {{v}^2} = - \gamma \frac{\left[\mbox{\boldmath$\beta$} \times
\mbox{\boldmath$a$}\right]} {c{\beta}^2} 
\label{Rel_Thomas_av_Jackson}
\end{equation}
for \(\beta\) approaching unity. (The relativistic version of the centrifugal force on which Eq. (\ref{AntiCentForceDef}) is based, that includes the leading Lorentz factor, \(\gamma\), is derived in \cite{ArendtY98}.) The anti-centrifugal force according to Eq. (\ref{AntiCentForceDef}) on one particle due to the other is thus,
\begin{eqnarray}
\mbox{\boldmath $F$}_{\text{a.c.}} =  \frac{\gamma_2 m_2} {c^2{\beta_1}^4}{\gamma_1}^2 \left[\mbox{\boldmath$\beta$}_1 \times
\mbox{\boldmath$a$}_1\right]\times \left(\left[\mbox{\boldmath$\beta$}_1 \times
\mbox{\boldmath$a$}_1\right] \times \mbox{\boldmath $r$}_{12}\right).
\label{centforce2}
\end{eqnarray}
where \(\mbox{\boldmath$r$}_{12} \equiv \mbox{\boldmath$r$}_{1} - \mbox{\boldmath$r$}_{2}\), and where the two particles are now distinguished by the subscripts 1 and 2, since the field-source particle is no longer constrained to be non-accelerating, and in fact is accelerating under the influence of the Coulomb field of particle 2. Therefore the notion of a test particle that doesn't influence the field-source particle motion must be abandoned, and the subscripts \(s\) and \(t\) have been replaced by \(1\) and \(2\).

For Coulomb attraction or repulsion in the case of mutual circular motion of the two charged particles and neglecting effects of retardation, Eq. (\ref{accel_from_RLOM}) gives the acceleration of one of the particles in the laboratory frame due to the velocity electric field according to the exact electric field expression of Eq. (\ref{LW_E_vel_field}) of the other as
\begin{equation}
\mbox{\boldmath$a$}_1 = \left[\frac{1}{\gamma_1 m_1} \right]\frac{q_1 q_2\mbox{\boldmath$r$}_{12}}{{\gamma_2}^2 R^3}. 
\label{CoulAccelRel}
\end{equation}

In the approximation valid in the highly relativistic case where \(\beta_1\) and \(\beta_2\) approach unity and for circular motion neglecting retardation, and assuming the acceleration of particle 1 is Coulombic (but retaining the inverse \({\gamma_1}^2\) factor  in the acceleration according the exact electric field expression of Eq. (\ref{LW_E_vel_field})), the expected magnitude of the anti-centrifugal force acting on particle 2 is thus
\begin{widetext}
\begin{eqnarray}
\big|\mbox{\boldmath $F$}_{\text{a.c.}}\big| \approx \frac{\gamma_2 m_2}{c^2}\big|\mbox{\boldmath$a$}_1\big|^2 R \approx  \frac{\gamma_2 m_2} {c^2}\frac{{q_1}^2{q_2}^2{\gamma_1}^2}{{\gamma_1}^2{m_1}^2 {\gamma_2}^4 R^3} = \frac{ m_2} {c^2}\frac{{q_1}^2{q_2}^2}{{m_1}^2 {\gamma_2}^3 R^3}.
\label{AntiCentForceCirc}
\end{eqnarray}
\end{widetext}

If it is assumed the two particles are of like charge polarity with  charge magnitude of the order of the elementary charge, and repulsed by Coulomb repulsion, yet circularly orbiting each other with laboratory-frame velocities near the speed of light, then it will be a simple matter to calculate approximately (\textit{i.e.}, neglecting retardation) at what orbital radius the anticentrifugal force will overcome Coulomb repulsion. This orbital radius can then be compared with the measured size of a nucleon such as the proton.

Equating the anticentrifugal force of Eq. (\ref{AntiCentForceCirc}) and the magnitude of the electric force based on the electric field of Eq. (\ref{LW_E_vel_field}) but neglecting retardation obtains, for equal mass particles and asumming \(|q_1|=|q_2| = e\) (with \(e\) the fundamental charge magnitude) and \(\gamma_1=\gamma_2 \equiv \gamma\), and using the proton mass for \({\gamma} m\),
\begin{equation}
R \approx \frac{e^2}{\gamma m c^2}\approx 10^{-15} \text{cm}.
\label{proton_rad_ll}
\end{equation}

This orbital radius or (strictly) inter-particle separation is about two orders of magnitude less than the estimated size of the proton \cite{MohrY11}.  Thus the present analysis would seem to indicate that the expected anticentrifugal force of the Thomas precession is too weak to potentially account for the fundamental strong force that binds quarks into nucleons.  However, it would seem a valid candidate for the needed force that could bind conjectured constituent particles of quarks and leptons \cite{LincolnY12}.  Furthermore, it is plausible that a force of relativistic kinematical origin such as the anti-centrifugal force of the Thomas precession would not have the same binding energy and relativistic mass equivalency usually expected, as needed to bind such constituent particles ({\em e.g.,} ``preons'') without exceeding the known masses of the leptons and quarks. 

Although the Thomas precession is widely credited with accounting for the spin-orbit coupling being half its otherwise classically expected value, the original analysis \cite{Thomas1927} was performed prior to recognition of the need to account for the ``hidden'' relativistic momentum of a magnetic dipole in an electric field.  Accounting for hidden momentum can be seen to directly affect this conclusion \cite{LushY9}.  Perhaps further re-examination of the relationship of the Thomas precession to energy is warranted.

\section{The Strong Force of Maxwell-Lorentz Electrodynamics}

It is worth noting that a force similar to the expected anti-centrifugal force of the Thomas precession can be found in relativistic electrodynamics. If retardation effects can be neglected, conventional electrodynamics predicts that two charges closely approaching each other at relativistic relative velocity experience a mutually attractive force that is independendent of relative polarity and similar in strength to the predicted anti-centrifugal force due to Thomas precession.  

Based on the Li\'enard-Wiechert field expressions provided above as Eqs. (\ref{LW_E_field}) and (\ref{LW_B_field}), the magnetic force on particle 2 due to the acceleration field of particle 1 is generally
\begin{widetext}
\begin{equation}
\mbox{\boldmath$F$}_m = \frac{q_2 \mbox{\boldmath$v$}_2}{c} \times \mbox{\boldmath$B$} = q_2 \mbox{\boldmath$\beta$}_2 \times \left[ \mbox{\boldmath$n$} \times \left[\frac{q_1}{c} \left[ \frac{\mbox{\boldmath$n$} \times \left((\mbox{\boldmath$n$}-\mbox{\boldmath$\beta$}_1)\times \dot{\mbox{\boldmath$\beta$}}_1\right)}{\left(1 - \mbox{\boldmath$\beta$}_1 \cdot \mbox{\boldmath$n$}\right)^3 R }\right]\right] \right]_{\text{ret}},
\end{equation}
\end{widetext}
where here \( \mbox{\boldmath$n$} \equiv \mbox{\boldmath$r$}_{21}/R \equiv (\mbox{\boldmath$r$}_2 - \mbox{\boldmath$r$}_1)/R\)  with \(R \equiv |\mbox{\boldmath$r$}_2 - \mbox{\boldmath$r$}_1|\).  The acceleration of particle 1 in the electric velocity field of particle 2 is found from Eqs. (\ref{accel_from_RLOM}) and (\ref{LW_E_field}), neglecting retardation and assuming mutually circular orbital motion of particle 1 around particle 2, as  
\begin{equation}
\mbox{\boldmath$a$}_1 \approx \frac{ q_2 q_1 }{\gamma_1 m_1 {\gamma_2}^2 R^3}\left[\mbox{\boldmath$r$}_{12}\right] \equiv -\frac{ q_2 q_1 }{\gamma_1 m_1 {\gamma_2}^2 R^2}\left[\mbox{\boldmath$n$}\right],
\end{equation}
or, if \( \dot{\mbox{\boldmath$\beta$}}_1 = \mbox{\boldmath$a$}_1/c \approx -a_1\mbox{\boldmath$n$}/c\) with \(a_1 \equiv |\mbox{\boldmath$a$}_1|\), and still neglecting retardation,
\begin{equation}
\mbox{\boldmath$F$}_{\text{magnetic}} \approx    \mbox{\boldmath$\beta$}_2 \times \left[ \mbox{\boldmath$n$} \times \left[  \mbox{\boldmath$n$} \times \left(\mbox{\boldmath$\beta$}_1\times \mbox{\boldmath$n$}\right)\right]\right]\frac{q_2 q_1}{{c^2 R}}\frac{q_2 q_1}{\gamma_1 m_1 {\gamma_2}^2 R^2}.
\label{StrongMagneticForce}
\end{equation}

Now assume mutual circular motion and let \(\hat{\mbox{\boldmath$x$}},\hat{\mbox{\boldmath$y$}},\hat{\mbox{\boldmath$z$}}\) represent orthogonal unit vectors in a right-handed coordinate system, and align \(\hat{\mbox{\boldmath$x$}}\) with \(\mbox{\boldmath$n$} \) and \(\hat{\mbox{\boldmath$y$}}\) with \(\mbox{\boldmath$v$}_1\).  Then for circular motion with \(\mbox{\boldmath$\beta$}_2 = -\mbox{\boldmath$\beta$}_1 \) and \(|\mbox{\boldmath$\beta$}_2| = |\mbox{\boldmath$\beta$}_1| \approx 1 \), and neglecting retardation, Eq. (\ref{StrongMagneticForce}) for the magnetic force on particle 2 due to its motion in the magnetic acceleration field of particle 1 becomes 
\begin{equation}
\mbox{\boldmath$F$}_{\text{magnetic}} \approx (-\mbox{\boldmath$n$}) \frac{{q_2}^2 {q_1}^2}{\gamma_1 m_1 c^2 {\gamma_2}^2 R^3}.
\label{StrongMagneticForceCirc}
\end{equation}

Recalling that \(\mbox{\boldmath$n$}\) is directed towards particle 2 from particle 1, it is apparent that this force is attractive for like charges, as well as for opposite.  Also, comparing Eq. (\ref{StrongMagneticForceCirc}) with Eq. (\ref{AntiCentForceCirc}), it is apparent that, for mutual circular motion and neglecting retardation,
\begin{equation}
\mbox{\boldmath$F$}_{\text{magnetic}} \approx \frac{{\gamma_2} m_1}{{\gamma_1} m_2} \mbox{\boldmath $F$}_{\text{anticentrifugal}}.  
\end{equation}

In the case of equal mass particles with \(\gamma_1 = \gamma_2\) so that \(\mbox{\boldmath$F$}_{\text{magnetic}} \approx  \mbox{\boldmath $F$}_{\text{anticentrifugal}}\), it is apparent that under  conditions as stated the magnetic force is approximately equal to the predicted anti-centrifugal force of the Thomas precession.

\section{Concluding Remarks}

It has been proposed that relativistic kinematic considerations necessitate that the Thomas precession gives rise to certain forces that resemble the rotational pseudoforces known as the Coriolis, Euler, and centrifugal forces.  It has been shown to order \(v^2/c^2\) that the predicted anti-Coriolis force of the Thomas precession accounts for the existence of the magnetic force. The possible correspondence of the strong nuclear force to an anti-centrifugal force of the Thomas procession was investigated.


\appendix

\section{Two-Particle Interaction Seen by an Observer Co-Moving with the Test Particle}

An observer co-moving with the test particle sees a Thomas precession of the field-source particle rest frame.  Also, in the source particle rest frame there is no magnetic force acting on the test particle, only the velocity-independent Coulomb force.   

At a succession of instants of time, an observer co-moving with the source particle can calculate the field source particle position relative to the test particle, as the source particle relative position would be expressed in the rest frame of the test particle.  

\subsection{Change of Field Source Particle Position as Measured by an Observer Co-Moving with the Test Particle}

Let \({\mbox{\boldmath$\eta$}}(\zeta), {\mbox{\boldmath$\mu$}}(\zeta) \)  represent the test particle position and velocity measured by an observer in the source particle rest frame (an inertial reference frame, since the field source particle is restricted in the present analysis to moving uniformly with respect to the inertial laboratory reference frame), as a function of time, \(\zeta\), in the source particle rest frame, and \({\mbox{\boldmath$\rho$}_s}(\tau), {\mbox{\boldmath$\nu$}_s}(\tau), {\mbox{\boldmath$\alpha$}_s}(\tau)\)  represent the source particle position, velocity, and acceleration relative to the test particle, as measured by an observer co-moving with the test particle, as a function of the test particle proper time \(\tau\).

Also let \({\mbox{\boldmath$\sigma$}_s}(\xi), {\mbox{\boldmath$\lambda$}_s}(\xi)\) represent the source particle position and velocity in an inertial reference frame momentarily co-moving with the test particle at time \(\zeta=\zeta_0\), where \(\xi\) is the source particle time coordinate in this inertial reference frame, which will be referred to as the test particle momentary rest frame. The space origin of the field source particle rest frame coordinate system is at the field source particle, and the observer co-moving and co-located with the test particle chooses a coordinate system for the inertial reference frame moving with the test particle at its proper time \(\tau=\tau_0\) and space origin at the test particle.  

Similarly, let \({\mbox{\boldmath$\eta$}_p}(\zeta), {\mbox{\boldmath$\mu$}_p}(\zeta) \) and  \({\mbox{\boldmath$\sigma$}_p}(\xi), {\mbox{\boldmath$\lambda$}_p}(\xi)\) represent the position and velocity of an arbitrarily located particle in the source particle rest frame and test particle momentary rest frame coordinate systems, respectively. 

The arbitrary particle position in the test particle momentary rest frame can be expressed accurately to order \(v^2/c^2\) (see Appendix D) in terms of test particle momentary rest frame at \(\zeta=\zeta_0\) quantities as 
\begin{equation}
{\mbox{\boldmath$\sigma$}_p}(\zeta) \approx \mbox{\boldmath$\eta$}_p -\mbox{\boldmath$\eta$}_0 + (\mbox{\boldmath$\beta$}_0 \cdot (\mbox{\boldmath$\eta$}_p -\mbox{\boldmath$\eta$}_0))  \mbox{\boldmath$\beta$}_0/2 - (\zeta-\zeta_0) \mbox{\boldmath$\mu$}_0 
\nonumber
\end{equation}
with \(\mbox{\boldmath$\beta$} \equiv \mbox{\boldmath$\mu$}/c \), and \(\gamma \equiv (1-(\mu/c)^2)^{-1/2}\), and \(\mbox{\boldmath$\beta$}_0 \equiv \mbox{\boldmath$\beta$}(t=t_0) \), and \(\gamma_0 \equiv \gamma(\zeta=\zeta_0)\). In the case that the arbitrary particle is the source particle, which is at the origin of its own rest frame here, then
\begin{equation}
{\mbox{\boldmath$\sigma$}_s}(\zeta) \approx -\mbox{\boldmath$\eta$}_0 - (\mbox{\boldmath$\beta$}_0 \cdot \mbox{\boldmath$\eta$}_0) \mbox{\boldmath$\beta$}_0/2 - (\zeta-\zeta_0) \mbox{\boldmath$\mu$}_0. 
\nonumber
\end{equation}

The space origin of the system \((\mbox{\boldmath$\sigma$},\xi) \) coincides with the test particle at \(\tau=\tau_0\), which corresponds to \(\xi=0\). Then the arbitrary particle time coordinate according to the observer co-moving with the test particle i
\begin{equation}
\xi(\zeta) =  (\zeta-\zeta_0)\gamma_0  - \gamma_0 \mbox{\boldmath$\beta$}_0 \cdot (\mbox{\boldmath$\eta$}_p(\zeta) -\mbox{\boldmath$\eta$}_0) / c. 
\nonumber
\end{equation}

Thus
\begin{equation}
\frac{d\xi}{d\zeta} =  \gamma_0 - \gamma_0 \mbox{\boldmath$\beta$}_0 \cdot \mbox{\boldmath$\mu$}_p(\zeta) / c. 
\nonumber
\end{equation}

Since the source particle is by definition stationary in the source particle rest frame, for the field source particle
\begin{equation}
\frac{d\xi}{d\zeta} =  \gamma_0,   
\end{equation}
and for the test particle
\begin{equation}
\frac{d\xi}{d\zeta}\big|_{(\zeta=\zeta_0)} =  \gamma_0 - \gamma_0 \mbox{\boldmath$\beta$}_0 \cdot \mbox{\boldmath$\mu$}(\zeta_0) / c \approx {\gamma_0}^{-1}.   
\end{equation}

Also, let 
\begin{equation}
\mbox{\boldmath$\sigma$}_{s0} \equiv \mbox{\boldmath$\sigma$}_s(\zeta=\zeta_0) 
 \approx -\mbox{\boldmath$\eta$}_0 - (\mbox{\boldmath$\beta$}_0 \cdot \mbox{\boldmath$\eta$}_0)  \mbox{\boldmath$\beta$}_0/2 
\nonumber
\end{equation}
with \(\mbox{\boldmath$\eta$}_0 \equiv \mbox{\boldmath$\eta$}(\zeta=\zeta_0)\), and
\begin{equation}
\xi_{0} \equiv \xi(\zeta=\zeta_0) = \delta \zeta \gamma_0 - \gamma_0 \mbox{\boldmath$\beta$}_0 \cdot (\mbox{\boldmath$\eta$}_{\delta} - \mbox{\boldmath$\eta$}_0) / c. 
\nonumber
\end{equation}

For the source particle this becomes
\begin{equation}
\xi_{0} \equiv \xi(\zeta=\zeta_0) = \gamma_0 \mbox{\boldmath$\beta$}_0 \cdot \mbox{\boldmath$\eta$}_0/c. 
\end{equation}

The arbitrary particle position in an inertial coordinate system momentarily co-moving with the test particle at time \(\zeta_0 +\delta \zeta\) can be expressed (accurately to order \(v^2/c^2\)) as
\begin{equation}
{\mbox{\boldmath$\sigma$}_p}' \approx \mbox{\boldmath$\eta$}_p -\mbox{\boldmath$\eta$}_\delta + (\mbox{\boldmath$\beta$}_\delta \cdot (\mbox{\boldmath$\eta$}_p - \mbox{\boldmath$\eta$}_\delta))  \mbox{\boldmath$\beta$}_\delta/2 - (\zeta - (\zeta_0 +\delta \zeta)) \mbox{\boldmath$\mu$}_\delta    
\nonumber
\end{equation}
with \(\mbox{\boldmath$\eta$}_\delta \equiv \mbox{\boldmath$\eta$}(\zeta=\zeta_0 + \delta \zeta)\), and with \(\mbox{\boldmath$\beta$}_\delta \equiv \mbox{\boldmath$\beta$}(\zeta=\zeta_0 + \delta \zeta)\), and \(\gamma_\delta \equiv \gamma(\zeta=\zeta_0 + \delta \zeta)\).  For the source particle this becomes 
\begin{equation}
{\mbox{\boldmath$\sigma$}_s}' \approx  -\mbox{\boldmath$\eta$}_\delta - (\mbox{\boldmath$\beta$}_\delta \cdot  \mbox{\boldmath$\eta$}_\delta) \mbox{\boldmath$\beta$}_\delta/2 - (\zeta - (\zeta_0 +\delta \zeta)) \mbox{\boldmath$\mu$}_\delta.    
\end{equation}

Similarly, let
\begin{equation}
{\xi}' =  (\zeta - (\zeta_0 + \delta \zeta)) \gamma_\delta - \gamma_\delta \mbox{\boldmath$\beta$}_\delta \cdot (\mbox{\boldmath$\eta$}_{0}-\mbox{\boldmath$\eta$}_\delta) / c. 
\end{equation}
Then
\begin{widetext}
\begin{equation}
{\mbox{\boldmath$\sigma$}}_{s\delta} \equiv {\mbox{\boldmath$\sigma$}_s}'(\zeta'= {\zeta'}_0)  \equiv {\mbox{\boldmath$\sigma$}_s}'(\zeta = \zeta_0 + \delta \zeta) \approx   -\mbox{\boldmath$\eta$}_{\delta} - (\mbox{\boldmath$\beta$}_{\delta} \cdot \mbox{\boldmath$\eta$}_{\delta}) \mbox{\boldmath$\beta$}_{\delta}/2 
\end{equation}
where \(\mbox{\boldmath$\eta$}_\delta  \equiv \mbox{\boldmath$\eta$}(\zeta=\zeta_0 + \delta \zeta)\), and
\begin{equation}
\xi_{\delta} \equiv {\xi}'(\zeta' = {\zeta'}_0) \equiv {\xi}'(\zeta = \zeta_0 + \delta \zeta) =  -\gamma_{\delta} \mbox{\boldmath$\beta$}_{\delta} \cdot (\mbox{\boldmath$\eta$}_{p\delta} - \mbox{\boldmath$\eta$}_{\delta})/c 
\end{equation}
or, in the case of the arbitrary particle being the field source particle so that \(\mbox{\boldmath$\eta$}_{p\delta} = \mbox{\boldmath$\eta$}_{s\delta} \equiv 0\),
\begin{equation}
\xi_{\delta} \equiv  {\xi}'(\zeta=\zeta_0 + \delta \zeta) =  \gamma_{\delta} \mbox{\boldmath$\beta$}_{\delta} \cdot  \mbox{\boldmath$\eta$}_{\delta} / c 
\end{equation}
with
\begin{equation}
\mbox{\boldmath$\eta$}_{\delta} \equiv \mbox{\boldmath$\eta$}(\zeta = \zeta_0 + \delta \zeta) \approx \mbox{\boldmath$\eta$}_0 + \delta \zeta \mbox{\boldmath$\mu$}_0
\end{equation}
where \( \mbox{\boldmath$\mu$} \) is the test particle velocity relative to the source particle (measured by a source particle co-moving observer). 

Now let \( \delta \mbox{\boldmath$\rho$}_s \equiv {\mbox{\boldmath$\sigma$}_s}'(\zeta_0 + \delta \zeta) - \mbox{\boldmath$\sigma$}(\zeta_0) \equiv \mbox{\boldmath$\sigma$}_{s\delta} - \mbox{\boldmath$\sigma$}_{s0} \). Then, 
\begin{eqnarray}
\delta \mbox{\boldmath$\rho$}_s \approx -\mbox{\boldmath$\eta$}_{\delta} - (\mbox{\boldmath$\beta$}_\delta \cdot \mbox{\boldmath$\eta$}_{\delta})  \mbox{\boldmath$\beta$}_\delta/2 + \mbox{\boldmath$\eta$}_0 + (\mbox{\boldmath$\beta$}_0 \cdot \mbox{\boldmath$\eta$}_0)  \mbox{\boldmath$\beta$}_0/2.  
\nonumber
\end{eqnarray}

Replacing \(\mbox{\boldmath$\eta$}_{\delta} \) by \(\mbox{\boldmath$\eta$}_{0} + \delta \zeta \mbox{\boldmath$\mu$}_{0} \) and \( \mbox{\boldmath$\beta$}_\delta\) by \(\mbox{\boldmath$\beta$}_0 + \delta \zeta \dot{\mbox{\boldmath$\beta$}}_0 \) and reducing obtains
\begin{eqnarray}
\delta \mbox{\boldmath$\rho$}_s \approx  -\delta \zeta \mbox{\boldmath$\mu$} - \delta \zeta \left(\mbox{\boldmath$\beta$} \cdot \mbox{\boldmath$\eta$} \right)\dot{\mbox{\boldmath$\beta$}}/2 - \delta \zeta\left(\dot{\mbox{\boldmath$\beta$}} \cdot \mbox{\boldmath$\eta$} \right)\mbox{\boldmath$\beta$} /2   - \delta \zeta\left(\mbox{\boldmath$\beta$} \cdot \mbox{\boldmath$\mu$} \right)\mbox{\boldmath$\beta$}/2 
\nonumber
\end{eqnarray}
so
\begin{eqnarray}
\lim_{\delta \zeta \rightarrow 0}\frac{\delta \mbox{\boldmath$\rho$}_s}{\delta \zeta} \equiv \frac{d\mbox{\boldmath$\rho$}_s}{d\zeta} \approx   -\mbox{\boldmath$\mu$} - \left(\mbox{\boldmath$\beta$} \cdot \mbox{\boldmath$\eta$} \right)  \dot{\mbox{\boldmath$\beta$}}/2 - \left( \dot{\mbox{\boldmath$\beta$}} \cdot \mbox{\boldmath$\eta$} \right)  \mbox{\boldmath$\beta$}/2  - \left(\mbox{\boldmath$\beta$} \cdot \mbox{\boldmath$\mu$}\right) \mbox{\boldmath$\beta$}/2 
\nonumber
\end{eqnarray}
\end{widetext}
or, since \(\left(1 + \left(\mbox{\boldmath$\beta$} \cdot \mbox{\boldmath$\mu$}\right)/2 \right) \mbox{\boldmath$\beta$} = (1 + \beta^2/2)\mbox{\boldmath$\mu$} \approx \gamma \mbox{\boldmath$\mu$}\),
\begin{eqnarray}
\frac{d\mbox{\boldmath$\rho$}_s}{d\zeta}  \equiv \mbox{\boldmath$\nu$}_s \approx -\gamma\mbox{\boldmath$\mu$}  - \left(\mbox{\boldmath$\beta$} \cdot \mbox{\boldmath$\eta$} \right)  \dot{\mbox{\boldmath$\beta$}}/2 - \left( \dot{\mbox{\boldmath$\beta$}} \cdot \mbox{\boldmath$\eta$} \right)  \mbox{\boldmath$\beta$}/2  
\end{eqnarray}

\subsection{Rate of Change of Field Source Particle Time Coordinate as Measured by an Observer Co-Moving with the Test Particle}

It was found above that the arbitrarily located particle time coordinate in the test particle momentary rest frame with origin at the test particle position at time \(\zeta=\zeta_0\) in the field source particle rest frame, corresponding to time \(\zeta=\zeta_0\) in the field source particle rest frame, can be expressed as 
\begin{equation}
\xi_{0} \equiv \xi(\zeta=\zeta_0) = \delta \zeta \gamma_0 - \gamma_0 \mbox{\boldmath$\beta$}_0 \cdot (\mbox{\boldmath$\eta$}_{p\delta} - \mbox{\boldmath$\eta$}_0) / c. 
\nonumber
\end{equation}

For the source particle this becomes
\begin{equation}
\xi_{0} = \gamma_0 \mbox{\boldmath$\beta$}_0 \cdot \mbox{\boldmath$\eta$}_0/c. 
\label{source_time_1}
\end{equation}

Suppose temporarily that the test particle is not accelerating, and  suppose that at a slightly later time \(\zeta_0 + \delta \zeta\) we perform a boost so that
\begin{equation}
{\xi'}(\zeta) =  (\zeta-(\zeta_0 + \delta \zeta ))\gamma_0  - \gamma_0 \mbox{\boldmath$\beta$}_0 \cdot (\mbox{\boldmath$\eta$}_p(\zeta + \delta \zeta) - \mbox{\boldmath$\eta$}_\delta)/c. 
\nonumber
\end{equation}

For the arbitrary particle being the source particle, which is staionary at the origin of its rest frame coordinates, this becomes
\begin{equation}
{\xi'}(\zeta) =  (\zeta-(\zeta_0 + \delta \zeta ))\gamma_0  + \gamma_0 \mbox{\boldmath$\beta$}_0 \cdot \mbox{\boldmath$\eta$}_\delta / c. 
\end{equation}

At \(\zeta = \zeta_0 + \delta \zeta\),
\begin{equation}
{\xi'}(\zeta_0 + \delta \zeta) =  \gamma_0 \mbox{\boldmath$\beta$}_0 \cdot \mbox{\boldmath$\eta$}_\delta / c. 
\label{source_time_2}
\end{equation}

In the above analysis, the change of the source particle position was found between two inertial coordinate systems with space origins at the test particle, and where the accelerating test particle is momentarily at rest, at two successive times separated by a small time interval of \(\delta \zeta_0\) in the source particle rest frame. In order to determine a rate of change of the source particle position relative to the test particle as measured by the observer co-moving with the test particle, it is necessary to determine the corresponding interval of time that elapses between the two successive positions, according to the test particle co-moving observer.  However, in the limiting case that the test particle is  not accelerating, the difference between the two time coordinates given by Eqs. (\ref{source_time_1}) and (\ref{source_time_2}) vanishes.  It is thus needed to find the translation of  time that will place the time of the successive snapshots of position consistently with the elapsed test particle proper time.  That is, it is necessary to find a time translation \(\Delta\) and a translated time coordinate  \({\xi''}(\zeta) = {\xi'}(\zeta) + \Delta\), such that \({\xi''}(\zeta_0 + \delta \zeta) = \xi(\zeta_0 + \delta \zeta) = \gamma_0 \delta \zeta \equiv \delta \tau\), for the case of a non-accelerating test particle.  So, let
\begin{equation}
\Delta = {\xi}(\zeta_0 + \delta \zeta) - {\xi'}(\zeta_0 + \delta \zeta)  
\end{equation}
or
\begin{equation}
\Delta = \delta \zeta \gamma_0 + \gamma_0 \mbox{\boldmath$\beta$}_0 \cdot \mbox{\boldmath$\eta$}_0 / c - \gamma_0 \mbox{\boldmath$\beta$}_0 \cdot \mbox{\boldmath$\eta$}_\delta / c. 
\end{equation}

Substitution for \(\mbox{\boldmath$\eta$}_\delta \approx \mbox{\boldmath$\eta$}_0 + \delta \zeta \mbox{\boldmath$\mu$}_0 \) and reducing obtains that
\begin{equation}
\Delta \approx \delta \zeta (\gamma_0 - \mbox{\boldmath$\beta$}_0 \cdot  \mbox{\boldmath$\mu$}_0 / c) 
\end{equation}
and so
\begin{equation}
{\xi''}(\zeta) \approx  (\zeta-\zeta_0 )\gamma_0  + \mbox{\boldmath$\beta$}_0 \cdot \mbox{\boldmath$\eta$}_0 / c 
\end{equation}
which becomes, for the case of an accelerating test particle
\begin{equation}
{\xi''}(\zeta) \approx  (\zeta-\zeta_0 )\gamma_\delta  + \mbox{\boldmath$\beta$}_\delta \cdot \mbox{\boldmath$\eta$}_0 / c. 
\end{equation}

Allowing for test particle acceleration, let
\begin{equation}
\delta \tau \equiv {\xi''}(\zeta_0 + \delta \zeta) - {\xi}_s(\zeta_0) \approx  \delta \zeta \gamma_\delta  + \mbox{\boldmath$\beta$}_\delta \cdot \mbox{\boldmath$\eta$}_0 / c -  \mbox{\boldmath$\beta$}_0 \cdot \mbox{\boldmath$\eta$}_0 / c
\nonumber
\end{equation}
or
\begin{equation}
\delta \tau \approx  \delta \zeta \gamma_0  + \delta \zeta \dot{\mbox{\boldmath$\beta$}} \cdot \mbox{\boldmath$\eta$}_0 / c
\end{equation}
or (dropping the subscript zero because the time \(\zeta_0\) is arbitrary),
\begin{equation}
\delta \tau \approx  \delta \zeta \gamma  + \delta \zeta \dot{\mbox{\boldmath$\beta$}} \cdot \mbox{\boldmath$\eta$} / c 
\end{equation}
so
\begin{equation}
\lim_{\delta \zeta \rightarrow 0}\frac{\delta \tau}{\delta \zeta} \equiv \frac{d \tau}{d \zeta} \approx   \gamma  +  \dot{\mbox{\boldmath$\beta$}} \cdot \mbox{\boldmath$\eta$} / c 
\end{equation}
and 
\begin{equation}
\frac{d \zeta} {d \tau} \approx   \gamma^{-1} - \dot{\mbox{\boldmath$\beta$}} \cdot \mbox{\boldmath$\eta$}/c.
\end{equation}

\subsection{Rate of Change of Field Source Particle Position as Measured by an Observer Co-Moving with the Test Particle}

The source particle velocity as measured by the test particle co-moving observer can be evaluated as 
\begin{eqnarray}
\mbox{\boldmath$\nu$}_s \equiv \frac{d{\mbox{\boldmath$\rho$}_s}}{d\tau}    = \frac{d{\mbox{\boldmath$\rho$}_s}}{d\zeta} \frac{d\zeta}{d\tau}. 
\end{eqnarray}

Using results from above, then, 
\begin{widetext}
\begin{eqnarray}
\mbox{\boldmath$\nu$}_s \approx -\left[\gamma  \mbox{\boldmath$\mu$} + \left(\dot{\mbox{\boldmath$\beta$}} \cdot \mbox{\boldmath$\eta$}\right) \mbox{\boldmath$\beta$} /2  +  \left(\mbox{\boldmath$\beta$} \cdot \mbox{\boldmath$\eta$}\right) \dot{\mbox{\boldmath$\beta$}}/2   \right]\left[ \gamma^{-1}  -  \dot{\mbox{\boldmath$\beta$}} \cdot \mbox{\boldmath$\eta$} / c  \right]
\end{eqnarray}
\end{widetext}
or
\begin{eqnarray}
\mbox{\boldmath$\nu$}_s \approx  -\mbox{\boldmath$\mu$} + \left(\dot{\mbox{\boldmath$\beta$}} \cdot \mbox{\boldmath$\eta$}\right) \mbox{\boldmath$\beta$} /2  -  \left(\mbox{\boldmath$\beta$} \cdot \mbox{\boldmath$\eta$}\right) \dot{\mbox{\boldmath$\beta$}}/2.  
\end{eqnarray}

If the test particle rest frame is taken as the non-rotating frame, then the (relative) velocity in that frame is expected to be expressible as 
\begin{eqnarray}
\frac{d{\mbox{\boldmath$\rho$}}}{d\tau} \equiv \mbox{\boldmath$\nu$} = \mbox{\boldmath$v$} + \mbox{\boldmath$\omega$} \times \mbox{\boldmath$r$}
\end{eqnarray}
with
\begin{eqnarray}
\mbox{\boldmath$\omega$} \times \mbox{\boldmath$r$} = -\left(\dot{\mbox{\boldmath$\beta$}} \cdot \mbox{\boldmath$r$}\right) \mbox{\boldmath$\beta$} /2 + \left(\mbox{\boldmath$\beta$} \cdot \mbox{\boldmath$r$}\right) \dot{\mbox{\boldmath$\beta$}}/2,  
\nonumber 
\end{eqnarray}
so with the vector identity \( \mbox{\boldmath$a$} \times (
\mbox{\boldmath$b$} \times \mbox{\boldmath$c$}) =  (\mbox{\boldmath$c$}\cdot 
\mbox{\boldmath$a$})  \mbox{\boldmath$b$} - (\mbox{\boldmath$b$}\cdot 
\mbox{\boldmath$a$} ) \mbox{\boldmath$c$} \), with \( \mbox{\boldmath$a$} = \mbox{\boldmath$r$}\),  \( \mbox{\boldmath$b$} = \mbox{\boldmath$\beta$}\),  \( \mbox{\boldmath$c$} = \dot{\mbox{\boldmath$\beta$}}\),
\begin{eqnarray}
\mbox{\boldmath$\omega$} \approx \left(\mbox{\boldmath$\beta$} \times  \dot{\mbox{\boldmath$\beta$}}\right) /2  = -\frac{\mbox{\boldmath$a$} \times  \mbox{\boldmath$v$}}{2c^2} = -\mbox{\boldmath$\omega$}_{\text{T}} \equiv \mbox{\boldmath$\omega$}'_{\text{T}}.
\end{eqnarray}

Comparing this result with Eq. (\ref{Thomas_av_approx}), it appears to be opposite in sign.  However, the sign difference is easily accounted for by noting that the accelerated frame is taken as non-rotating here.  As seen from the laboratory frame, the accelerated frame rotates oppositely.   That is, if  (Jackson \cite{jcksn:classelec} Eq. (11.107), for a general vector \(\mbox{\boldmath$G$}\) )
\begin{eqnarray}
\left(\frac{d{\mbox{\boldmath$G$}}}{dt}\right)_{\text{nonrot}} = \left(\frac{d{\mbox{\boldmath$G$}}}{dt}\right)_{\text{rest frame}} + \mbox{\boldmath$\omega$}_{\text{T}} \times \mbox{\boldmath$G$}
\nonumber 
\end{eqnarray}
or, more generally,
\begin{eqnarray}
\left(\frac{d{\mbox{\boldmath$G$}}}{dt}\right)_{\text{nonrot}} = \left(\frac{d{\mbox{\boldmath$G$}}}{dt}\right)_{\text{rot}} + \mbox{\boldmath$\omega$} \times \mbox{\boldmath$G$}.
\nonumber 
\end{eqnarray}

So if
\begin{eqnarray}
\left(\frac{d{\mbox{\boldmath$G$}}}{dt}\right)_{\text{laboratory}} = \left(\frac{d{\mbox{\boldmath$G$}}}{dt}\right)_{\text{rest frame}} + \mbox{\boldmath$\omega$} \times \mbox{\boldmath$G$}
\nonumber 
\end{eqnarray}
then
\begin{eqnarray}
\left(\frac{d{\mbox{\boldmath$G$}}}{dt}\right)_{\text{rest frame}} = \left(\frac{d{\mbox{\boldmath$G$}}}{dt}\right)_{\text{laboratory}} + \left(-\mbox{\boldmath$\omega$}\right) \times \mbox{\boldmath$G$}.
\nonumber 
\end{eqnarray}

Interchanging the designation of which reference frame is taken as the rotating frame thus inverts the sign of the associated angular velocity.

\section{The Magnetic Force as a Purely Relativistic-Kinematic Effect of Coulomb Acceleration}

In  this section, it is shown explicitly that presence of a magnetic force on a test charge involves necessarily a Coulombic acceleration of the test particle, in at least one inertial reference frame. The argument is based on the interaction of two charged particles, but because one of them is constrained to be non-accelerating, the argument extends by the principle of linear superposition to the case of a magnetic field generated by a current-carrying neutral wire. If, in addition to relatively accelerating, the field-source particle and the test particle are also relatively translating with a component transverse to the acceleration, the field-source particle rest frame is then necessarily Thomas precessing with respect to an observer co-moving with the test particle.  This is precisely the condition under which a magnetic force is seen to act on the test particle by an observer in the laboratory frame. Therefore, the magnetic force can always be associated with Thomas precession of the source particle rest frame from the point of view of an observer co-moving with the test particle.

\subsection{Implications of the Relativistic Law of Inertia}

The relativistic law of inertia for the test particle in the field source particle rest frame (which is an inertial frame here since the source particle is non-accelerating) may be written as 
\begin{equation}
\left(\frac{d\left[\gamma_t \mbox{\boldmath $v$}_t\right]}{d\tau}\right)^{(\text{srf})} = \frac{\mbox{\boldmath $F$}^{(\text{srf})}}{m_t}, 
\label{SRFLoM}
\end{equation}
where the parenthetic superscript (srf) indicates quantities defined in the field-source particle rest frame. The demonstration will proceed by rewriting Eq. (\ref{SRFLoM}) in terms of laboratory frame quantities.  To facilitate evaluation of the left hand side of Eq. (\ref{SRFLoM}) in terms of lab-frame quanitities, consider (based on the Lorentz transformation as given in Appendix B below) that the time component of the four-vector displacement from the source to the test particle Lorentz transforms from a 4-coordinate system in the laboratory reference frame to one in the field source particle frame, where the origins of both systems are the same event, as
\begin{equation}
{c t}^{(\text{srf})} \equiv c\tau = \gamma_s  ( c t - \beta_s(\hat{\mbox{\boldmath$v$}}_s \cdot \mbox{\boldmath$r$}_t))
\end{equation}
where \(t\) is the time coordinate, in the laboratory frame, of the test particle when it is at the position \(\mbox{\boldmath$r$}_t\). \(\mbox{\boldmath$v$}_s\) is the field-source particle velocity in the laboratory frame.  Thus
\begin{equation}
\tau = \gamma_s  (t - (\mbox{\boldmath$\beta$}_s \cdot \mbox{\boldmath$r$}_t)/c), 
\end{equation}
and so (and with the source particle non-accelerating)
\begin{equation}
\frac{d\tau}{dt} =  \gamma_s(1 - \mbox{\boldmath$\beta$}_s \cdot  \dot{\mbox{\boldmath$r$}}_t/c) = \gamma_s(1 - \mbox{\boldmath$\beta$}_s \cdot  \mbox{\boldmath$\beta$}_t).
\end{equation}

For the field source particle moving at constant velocity, then, 
\begin{equation}
\frac{d\left(\mbox{\boldmath$P$}^{(\text{srf})}\right)}{dt} = \frac{d\left(\mbox{\boldmath$P$}^{(\text{srf})}\right)}{d\tau}\frac{d\tau}{dt} = \frac{d\left(\mbox{\boldmath$P$}^{(\text{srf})}\right)}{d\tau} \left[\gamma_s(1 - \mbox{\boldmath$\beta$}_s \cdot  \mbox{\boldmath$\beta$}_t)  \right].
\end{equation}

Thus,
\begin{equation}
\frac{d\left(\mbox{\boldmath$P$}^{(\text{srf})}\right)}{d\tau} =  \frac{d\left(\mbox{\boldmath$P$}^{(\text{srf})}\right)}{dt} \left[\gamma_s(1 - \mbox{\boldmath$\beta$}_s \cdot  \mbox{\boldmath$\beta$}_t) \right]^{-1}, 
\end{equation}
and so 
\begin{equation}
\frac{d\left(\left[\gamma_t \mbox{\boldmath$v$}_t\right]^{(\text{srf})}\right)}{dt} = \frac{ \mbox{\boldmath$F$}^{(\text{srf})}}{m_t} \left[\gamma_s(1 - \mbox{\boldmath$\beta$}_s \cdot  \mbox{\boldmath$\beta$}_t)  \right].
\label{dPbydtau}
\end{equation}

Expanding the left hand side of Eq. (\ref{dPbydtau}) obtains
\begin{widetext}
\begin{equation}
\left(\left[\gamma_t\right]^{(\text{srf})}\right) \frac{d\left(\left[\mbox{\boldmath$v$}_t\right]^{(\text{srf})}\right)}{dt}  + \left(\left[\mbox{\boldmath$v$}_t\right]^{(\text{srf})}\right) \frac{d\left(\left[\gamma_t \right]^{(\text{srf})}\right)}{dt}  = \frac{\mbox{\boldmath$F$}^{(\text{srf})}}{ m_t}\left[\gamma_s(1 - \mbox{\boldmath$\beta$}_s \cdot  \mbox{\boldmath$\beta$}_t)  \right],
\label{EffectiveForce}
\end{equation}
where (to order \(\beta^2\))
\begin{equation}
(\mbox{\boldmath$v$}_t)^{(\text{srf})}  \approx (1 + \mbox{\boldmath$\beta$}_t \cdot \mbox{\boldmath$\beta$}_s) \mbox{\boldmath$v$} - ({\beta_s}^2/2) \mbox{\boldmath$v$}_t + (\mbox{\boldmath$\beta$}_s \cdot \mbox{\boldmath$v$}_t)\mbox{\boldmath$\beta$}_s/2, 
\label{Expandt1a}
\end{equation}
so (and with the source particle non-accelerating)
\begin{eqnarray}
\frac{d\left(\left[ \mbox{\boldmath$v$}_t\right]^{(\text{srf})}\right)}{dt} \approx   -({\beta_s}^2/2) \mbox{\boldmath$a$}_t + \left[\mbox{\boldmath$\beta$}_s \cdot \mbox{\boldmath$a$}_t\right]\mbox{\boldmath$\beta$}_s/2  + (\dot{\mbox{\boldmath$\beta$}}_t \cdot \mbox{\boldmath$\beta$}_s ) \mbox{\boldmath$v$} + (1 + \mbox{\boldmath$\beta$}_t \cdot \mbox{\boldmath$\beta$}_s) \mbox{\boldmath$a$}.
\label{dvtblfbydt}
\end{eqnarray}

Also,
\begin{equation}
(\gamma_t)^{(\text{srf})} =   \gamma_s\gamma_t \left(1 - \mbox{\boldmath$\beta$}_t \cdot \mbox{\boldmath$\beta$}_s \right),
\end{equation}
so, to order \(\beta^2\),
\begin{equation}
\left(\frac{d\gamma_t}{dt}\right)^{(\text{srf})} \approx   \left[ \gamma_s\dot{\gamma}_t \left(1 - \mbox{\boldmath$\beta$}_t \cdot \mbox{\boldmath$\beta$}_s \right) - \gamma_s\gamma_t \left(\dot{\mbox{\boldmath$\beta$}}_t \cdot \mbox{\boldmath$\beta$}_s  \right)\right].
\end{equation}

With \(\dot{\gamma} = \gamma^3 \mbox{\boldmath$\beta$} \cdot \dot{\mbox{\boldmath$\beta$}} \approx \mbox{\boldmath$\beta$} \cdot \dot{\mbox{\boldmath$\beta$}}\) and keeping only to order \(\beta^2\), 
\begin{equation}
\left(\frac{d\gamma_t}{dt}\right)^{(\text{srf})} \approx   \left[ \left(\mbox{\boldmath$\beta$}_t \cdot \dot{\mbox{\boldmath$\beta$}}_t  \right)  -  \left(\dot{\mbox{\boldmath$\beta$}}_t \cdot \mbox{\boldmath$\beta$}_s \right)\right]= \left(\mbox{\boldmath$\beta$} \cdot \dot{\mbox{\boldmath$\beta$}}_t  \right),
\end{equation}
so
\begin{equation}
(\gamma_t)^{(\text{srf})} \left(\frac{d\left[\mbox{\boldmath$v$}_t\right]}{dt}\right)^{(\text{srf})} +  \mbox{\boldmath $v$} \left( \mbox{\boldmath$\beta$} \cdot \dot{\mbox{\boldmath$\beta$}}_t \right) \approx \frac{\mbox{\boldmath$F$}^{(\text{srf})}}{ m_t}\left[\gamma_s(1 - \mbox{\boldmath$\beta$}_s \cdot  \mbox{\boldmath$\beta$}_t)  \right].
\end{equation}

Substituting using (\ref{dvtblfbydt}),
\begin{eqnarray}
(\gamma_t)^{(\text{srf})} \left( ({\beta_s}^2/2) \mbox{\boldmath$a$}_t + \left[(\mbox{\boldmath$\beta$}_s \cdot \mbox{\boldmath$a$}_t)\right]\mbox{\boldmath$\beta$}_s/2  + (\dot{\mbox{\boldmath$\beta$}}_t \cdot \mbox{\boldmath$\beta$}_s) \mbox{\boldmath$v$} + (1 + \mbox{\boldmath$\beta$}_t \cdot \mbox{\boldmath$\beta$}_s) \mbox{\boldmath$a$}\right) +  \mbox{\boldmath $v$} \left( \mbox{\boldmath$\beta$} \cdot \dot{\mbox{\boldmath$\beta$}} \right)  \approx \frac{\mbox{\boldmath$F$}^{(\text{srf})}}{ m_t}\left[\gamma_s(1 - \mbox{\boldmath$\beta$}_s \cdot  \mbox{\boldmath$\beta$}_t)  \right]
\end{eqnarray}
which reduces straightforwardly to
\begin{eqnarray}
-({\beta_s}^2/2) \mbox{\boldmath$a$}_t + \left[(\mbox{\boldmath$\beta$}_s \cdot \mbox{\boldmath$a$}_t)\right]\mbox{\boldmath$\beta$}_s/2    + (\gamma_t)^{(\text{srf})}  \mbox{\boldmath$a$}_t + (\mbox{\boldmath$\beta$}_t \cdot \mbox{\boldmath$\beta$}_s) \mbox{\boldmath$a$}  +  \mbox{\boldmath $v$} \left[ \left(\mbox{\boldmath$\beta$}_t \cdot \dot{\mbox{\boldmath$\beta$}}_t  \right) \right] \approx \frac{\gamma_s\mbox{\boldmath $F$}^{(\text{srf})}}{m_t}\left[1 - \mbox{\boldmath$\beta$}_s \cdot  \mbox{\boldmath$\beta$}_t \right],
\end{eqnarray}
(and since the field source particle is non-accelerating so that \(\mbox{\boldmath$a$} \equiv \mbox{\boldmath$a$}_t\)). Also have (from Eq. (\ref{accel_from_RLOM})) that
\begin{equation}
\mbox{\boldmath$a$}_t = \left[\frac{1}{\gamma_t m_t} \right]\left[\mbox{\boldmath$F$}_t - {\gamma_t}^3(\mbox{\boldmath$\beta$}_t \cdot\dot{\mbox{\boldmath$\beta$}}_t) m_t \mbox{\boldmath$v$}_t \right],
\label{E_accel}
\end{equation}
so
\begin{eqnarray}
-({\beta_s}^2/2) \mbox{\boldmath$a$}_t + \left[(\mbox{\boldmath$\beta$}_s \cdot \mbox{\boldmath$a$}_t)\right]\mbox{\boldmath$\beta$}_s/2  +  \left[\frac{(\gamma_t)^{(\text{srf})}}{\gamma_t m_t} \right]\left[\mbox{\boldmath$F$}_t - {\gamma_t}^3(\mbox{\boldmath$\beta$}_t \cdot\dot{\mbox{\boldmath$\beta$}}_t) m_t \mbox{\boldmath$v$}_t \right]+ (\mbox{\boldmath$\beta$}_t \cdot \mbox{\boldmath$\beta$}_s) \mbox{\boldmath$a$} \nonumber \\  +  \mbox{\boldmath$v$} \left[\left(\mbox{\boldmath$\beta$}_t \cdot \dot{\mbox{\boldmath$\beta$}}_t  \right) \right] \approx \frac{\gamma_s \mbox{\boldmath$F$}^{(\text{srf})}}{m_t} \left[1 - \mbox{\boldmath$\beta$}_s \cdot  \mbox{\boldmath$\beta$}_t \right],
\end{eqnarray}
or
\begin{eqnarray}
 -({\beta_s}^2/2) \mbox{\boldmath$a$}_t + \left[(\mbox{\boldmath$\beta$}_s \cdot \mbox{\boldmath$a$}_t)\right]\mbox{\boldmath$\beta$}_s/2  +  \left[\frac{(\gamma_t)^{(\text{srf})}}{\gamma_t m_t} \right]\mbox{\boldmath$F$}_t + (\mbox{\boldmath$\beta$}_t \cdot \mbox{\boldmath$\beta$}_s) \mbox{\boldmath$a$}   - \mbox{\boldmath $v$}_s \left[ \left(\mbox{\boldmath$\beta$}_t \cdot \dot{\mbox{\boldmath$\beta$}}_t  \right) \right]\approx \frac{\gamma_s\mbox{\boldmath$F$}^{(\text{srf})}}{m_t}\left[1 - \mbox{\boldmath$\beta$}_s \cdot  \mbox{\boldmath$\beta$}_t \right].
\end{eqnarray}

The electric field in the source rest frame due to the stationary source particle is
\begin{equation}
\mbox{\boldmath$E$}^{(\text{srf})} = q_s \left[ \frac{\mbox{\boldmath$n$}}{ {R}^2 }\right]^{(\text{srf})},
\label{LW_E_vel_field_at_s}
\end{equation}
so substituting \(q_t \mbox{\boldmath$E$}^{(\text{srf})}\) for \(\mbox{\boldmath$F$}^{(\text{srf})}\) obtains
\begin{eqnarray}
 - ({\beta_s}^2/2) \mbox{\boldmath$a$}_t + \left[(\mbox{\boldmath$\beta$}_s \cdot \mbox{\boldmath$a$}_t)\right]\mbox{\boldmath$\beta$}_s/2  +  \left[\frac{(\gamma_t)^{(\text{srf})}}{\gamma_t m_t} \right]\mbox{\boldmath$F$}_t + (\mbox{\boldmath$\beta$}_t \cdot \mbox{\boldmath$\beta$}_s) \mbox{\boldmath$a$}   - \mbox{\boldmath $v$}_s \left[\left(\mbox{\boldmath$\beta$}_t \cdot \dot{\mbox{\boldmath$\beta$}}_t  \right) \right]\approx \frac{\gamma_s}{ m_t} \left[\frac{q_s q_t \mbox{\boldmath$n$}}{ {R}^2 }\right]^{(\text{srf})} \left[1 - \mbox{\boldmath$\beta$}_s \cdot  \mbox{\boldmath$\beta$}_t \right].
\label{Eq46}
\end{eqnarray}

It is shown below that
\begin{equation}
\frac{{\mbox{\boldmath$n$}}^{(\text{srf})} }{\left[R^{(\text{srf})}\right]^2} =  \left[\frac{\mbox{\boldmath$n$} - {\gamma_s} \mbox{\boldmath$\beta$}_s - (1 - {\gamma_s}) (\hat{\mbox{\boldmath$v$}}_s \cdot \mbox{\boldmath$n$})  \hat{\mbox{\boldmath$v$}}_s }{ {\gamma_s}^3 R^2 \left[1  - (\mbox{\boldmath$\beta$}_s \cdot \mbox{\boldmath$n$}) \right]^{3} }   \right]_{\text{ret}}^{(\text{lab})},
\end{equation}
so, to order \({\beta_s}^2\),
\begin{equation}
\frac{{\mbox{\boldmath$n$}}^{(\text{srf})} }{\left[R^{(\text{srf})}\right]^2} \approx \left[\frac{\mbox{\boldmath$n$} - \mbox{\boldmath$\beta$}_s }{ {\gamma_s}^3 R^2 \left[1  - (\mbox{\boldmath$\beta$}_s \cdot \mbox{\boldmath$n$}) \right]^{3} }   \right]_{\text{ret}}^{(\text{lab})} + \left[\frac{(\mbox{\boldmath$\beta$}_s \cdot \mbox{\boldmath$n$})  \mbox{\boldmath$\beta$}_s/2 }{R^2} \right].
\label{Eq48}
\end{equation}

Substituting Eq. (\ref{Eq48}) into Eq. (\ref{Eq46}):
\begin{eqnarray}
 - ({\beta_s}^2/2) \mbox{\boldmath$a$}_t + \left[(\mbox{\boldmath$\beta$}_s \cdot \mbox{\boldmath$a$}_t)\right]\mbox{\boldmath$\beta$}_s/2  +  \left[\frac{(\gamma_t)^{(\text{srf})}}{\gamma_t m_t} \right]\mbox{\boldmath$F$}_t + (\mbox{\boldmath$\beta$}_t \cdot \mbox{\boldmath$\beta$}_s) \mbox{\boldmath$a$} \nonumber \\    - \mbox{\boldmath $v$}_s \left[\left(\mbox{\boldmath$\beta$}_t \cdot \dot{\mbox{\boldmath$\beta$}}_t  \right) \right]\approx \frac{\gamma_s q_s q_t}{m_t} \left[\left[\frac{\mbox{\boldmath$n$} - \mbox{\boldmath$\beta$}_s}{{\gamma_s}^3 R^2 \left[1  - (\mbox{\boldmath$\beta$}_s \cdot \mbox{\boldmath$n$}) \right]^{3}} \right]_{\text{ret}}^{(\text{lab})} + \left[\frac{(\mbox{\boldmath$\beta$}_s \cdot \mbox{\boldmath$n$}) \mbox{\boldmath$\beta$}_s/2 }{R^2} \right]\right]\left[1 - \mbox{\boldmath$\beta$}_s \cdot  \mbox{\boldmath$\beta$}_t \right],
\end{eqnarray}
or (suppressing notation indicating retardation and since all quantities are assumed defined in the laboratory frame unless indicated otherwise),
\begin{eqnarray}
 - ({\beta_s}^2/2) \mbox{\boldmath$a$}_t +  \left[\frac{(\gamma_t)^{(\text{srf})}}{\gamma_t m_t} \right]\mbox{\boldmath$F$}_t + (\mbox{\boldmath$\beta$}_t \cdot \mbox{\boldmath$\beta$}_s) \mbox{\boldmath$a$}       - \mbox{\boldmath $v$}_s \left[\left(\mbox{\boldmath$\beta$}_t \cdot \dot{\mbox{\boldmath$\beta$}}_t  \right) \right]\approx \frac{\gamma_s q_s q_t}{m_t} \left[\frac{\mbox{\boldmath$n$} - \mbox{\boldmath$\beta$}_s }{ {\gamma_s}^3 R^2 \left[1  - (\mbox{\boldmath$\beta$}_s \cdot \mbox{\boldmath$n$}) \right]^{3}} \right] \left[1 - \mbox{\boldmath$\beta$}_s \cdot  \mbox{\boldmath$\beta$}_t \right].
\end{eqnarray}

Substituting for \(
(\gamma_t)^{(\text{srf})} =   \gamma_s\gamma_t \left(1 - \mbox{\boldmath$\beta$}_t \cdot \mbox{\boldmath$\beta$}_s \right)
\), 
\begin{eqnarray}
- ({\beta_s}^2/2) \mbox{\boldmath$a$}_t  +  \left[\frac{\gamma_s \left(1 - \mbox{\boldmath$\beta$}_t \cdot \mbox{\boldmath$\beta$}_s \right)}{m_t} \right]\mbox{\boldmath$F$}_t + (\mbox{\boldmath$\beta$}_t \cdot \mbox{\boldmath$\beta$}_s) \mbox{\boldmath$a$}    - \mbox{\boldmath $v$}_s \left[ \left(\mbox{\boldmath$\beta$}_t \cdot \dot{\mbox{\boldmath$\beta$}}_t  \right) \right]\approx \frac{\gamma_s q_s q_t}{m_t} \left[\frac{\mbox{\boldmath$n$} - \mbox{\boldmath$\beta$}_s }{ {\gamma_s}^3 R^2 \left[1  - (\mbox{\boldmath$\beta$}_s \cdot \mbox{\boldmath$n$}) \right]^{3} }   \right] \left[1 - \mbox{\boldmath$\beta$}_s \cdot  \mbox{\boldmath$\beta$}_t \right].
\nonumber
\end{eqnarray}

With (for the non-accelerating field-source particle) \(\mbox{\boldmath$a$} \approx \mbox{\boldmath$F$}_t/m_t\),
\begin{eqnarray}
 - ({\beta_s}^2/2) \mbox{\boldmath$a$}_t  +  \left[\frac{\gamma_s}{m_t} \right]\mbox{\boldmath$F$}_t   - \mbox{\boldmath $v$}_s \left[ \left(\mbox{\boldmath$\beta$}_t \cdot \dot{\mbox{\boldmath$\beta$}}_t  \right) \right]\approx \frac{\gamma_s q_s q_t}{ m_t}\left[\frac{\mbox{\boldmath$n$} - \mbox{\boldmath$\beta$}_s }{ {\gamma_s}^3 R^2 \left[1  - (\mbox{\boldmath$\beta$}_s \cdot \mbox{\boldmath$n$}) \right]^{3} }   \right] \left[1 - \mbox{\boldmath$\beta$}_s \cdot  \mbox{\boldmath$\beta$}_t \right],
\end{eqnarray}
or (since \(\gamma_s \approx 1 + {\beta_s}^2/2\)),
\begin{eqnarray}
\left[\frac{1}{m_t} \right]\mbox{\boldmath$F$}_t   - \mbox{\boldmath $v$}_s \left[ \left(\mbox{\boldmath$\beta$}_t \cdot \dot{\mbox{\boldmath$\beta$}}_t  \right) \right]\approx \frac{\gamma_s q_s q_t}{ m_t}\left[\frac{\mbox{\boldmath$n$} - \mbox{\boldmath$\beta$}_s }{ {\gamma_s}^3 R^2 \left[1  - (\mbox{\boldmath$\beta$}_s \cdot \mbox{\boldmath$n$}) \right]^{3} } \right] \left[1 - \mbox{\boldmath$\beta$}_s \cdot  \mbox{\boldmath$\beta$}_t \right],
\end{eqnarray}
or
\begin{eqnarray}
\mbox{\boldmath$F$}_t  \approx \frac{q_s q_t}{ {\gamma_s}^2 } \left[\frac{\mbox{\boldmath$n$} - \mbox{\boldmath$\beta$}_s }{R^2 \left[1  - (\mbox{\boldmath$\beta$}_s \cdot \mbox{\boldmath$n$}) \right]^{3} } \right] - \left[\frac{q_s q_t}{R^2}   \right]\left[ (\mbox{\boldmath$\beta$}_s \cdot  \mbox{\boldmath$\beta$}_t)\mbox{\boldmath$n$} -   \left(\mbox{\boldmath$\beta$}_t \cdot \mbox{\boldmath$n$} \right)\mbox{\boldmath $\beta$}_s \right].
\end{eqnarray}
\end{widetext}

With the vector identity \(\mbox{\boldmath$a$}\times (
\mbox{\boldmath$b$} \times \mbox{\boldmath$c$}) =  (\mbox{\boldmath$c$}\cdot 
\mbox{\boldmath$a$})  \mbox{\boldmath$b$} - (\mbox{\boldmath$b$}\cdot 
\mbox{\boldmath$a$} ) \mbox{\boldmath$c$}\),
\begin{eqnarray}
\mbox{\boldmath$F$}_t  \approx \frac{q_s q_t}{ {\gamma_s}^2 } \left[\frac{\mbox{\boldmath$n$} - \mbox{\boldmath$\beta$}_s }{R^2 \left[1  - (\mbox{\boldmath$\beta$}_s \cdot \mbox{\boldmath$n$}) \right]^{3} } \right] - \left[\frac{q_s q_t}{R^2}   \right]\left[ \mbox{\boldmath$\beta$}_t \times (
\mbox{\boldmath$n$} \times \mbox{\boldmath$\beta$}_s) \right],
\nonumber
\end{eqnarray}
or
\begin{eqnarray}
\mbox{\boldmath$F$}_t  \approx \frac{q_s q_t}{ {\gamma_s}^2 } \left[\frac{\mbox{\boldmath$n$} - \mbox{\boldmath$\beta$}_s }{R^2 \left[1  - (\mbox{\boldmath$\beta$}_s \cdot \mbox{\boldmath$n$}) \right]^{3}} \right] + q_t \mbox{\boldmath$\beta$}_t \times \mbox{\boldmath$B$}, 
\end{eqnarray}
where
\begin{eqnarray}
\mbox{\boldmath$B$}  \approx \left[\frac{q_s}{R^2}   \right]\left[ 
\mbox{\boldmath$\beta$}_s  \times \mbox{\boldmath$n$} \right].
\end{eqnarray}

The above analysis demonstrates that the magnetic force can be interpreted as a relativistic-kinematic consequence of the acceleration of the test particle by the Coulomb force, and the relative motion of the field-source and test particles.

\section{Lorentz transformation, to the field source particle rest frame, of the field source particle to test particle null displacement, and the resulting electric field}

The Lorentz transformation for a general pure boost is \cite{jcksn:classelec}
\begin{widetext}
\begin{equation}
A_{\text{boost}}(\mbox{\boldmath$\beta$}) =  \left( \begin{array}{cccc} \gamma  & -\gamma \beta_1  & -\gamma \beta_2  &  -\gamma \beta_3  \\ -\gamma \beta_1  & 1 + \frac{(\gamma-1){\beta_1}^2}{\beta^2}  &  \frac{(\gamma-1)\beta_1\beta_2}{\beta^2}  &  \frac{(\gamma-1)\beta_1\beta_3}{\beta^2}  \\  -\gamma \beta_2  &  \frac{(\gamma-1)\beta_1\beta_2}{\beta^2}   & 1 + \frac{(\gamma-1){\beta_2}^2}{\beta^2}  &  \frac{(\gamma-1)\beta_2\beta_3}{\beta^2}  \\  -\gamma \beta_3   &  \frac{(\gamma-1)\beta_1\beta_3}{\beta^2}  &  \frac{(\gamma-1)\beta_2\beta_3}{\beta^2}  & 1 + \frac{(\gamma-1){\beta_3}^2}{\beta^2}  \end{array} \right)
\label{CartesianRotMetric}
\end{equation}

Suppose we identify the 1 direction as the direction of the test particle velocity in the field-source particle rest frame (where the field source particle is non-accelerating here), then
\begin{equation}
A_{\text{boost}}(\mbox{\boldmath$\beta$}) =  \left( \begin{array}{cccc} \gamma  & -\gamma \beta_1  & 0  &  0  \\ -\gamma \beta_1  & 1 + \frac{(\gamma-1){\beta_1}^2}{\beta^2}  & 0  &  0  \\  0  &  0   & 1   &  0 \\  0  &  0  &  0  & 1  \end{array} \right)=  \left( \begin{array}{cccc} \gamma  & -\gamma \beta_1  & 0  &  0  \\ -\gamma \beta_1  & \gamma  & 0  &  0  \\  0  &  0   &  1   &  0 \\  0  &  0  &  0  & 1  \end{array} \right)
\label{CartesianRotMetric2}
\end{equation}
where \(\beta_1 = \beta \) is the velocity of the boost here, and \(\gamma = 1/\sqrt{1-\beta^2}\).  So,

\end{widetext}
\begin{equation}
A_{\text{boost}}(\mbox{\boldmath$\beta$}) =  \left( \begin{array}{cccc} \gamma  & -\gamma \beta  & 0  &  0  \\ -\gamma \beta  & \gamma  & 0  &  0  \\  0  &  0   &  1   &  0 \\  0  &  0  &  0  & 1 \end{array} \right)
\label{CartesianRotMetric3}
\end{equation}

The four-position is generally 
\begin{equation}
{\cal R} \equiv (c t, \mbox{\boldmath$r$}), 
\end{equation}
where \(\mbox{\boldmath$r$} \) is the 3-vector position. The test particle four-position is then 	
\begin{equation}
{\cal R}_t = (c t, \mbox{\boldmath$r$}_t).
\end{equation}

The source particle four-position at the retarded time and space position of the source particle relative to the test particle position at time \(t\) is then 
\begin{equation}
{\cal R}_s = (c (t - R/c), \mbox{\boldmath$r$}_s(t - R/c)).
\end{equation}

The null four-displacement from the field source particle to the test particle  (to represent the field point being at the test particle) is then
\begin{equation}
{\cal R} \equiv {\cal R}_t - {\cal R}_s = (R, \mbox{\boldmath$r$}).
\end{equation}

The field source particle to test particle displacement can be put into components parallel and perpendicular to the test particle velocity \(\mbox{\boldmath$v$}\) relative to the source particle as
\begin{equation}
{\cal R}  =  (R,(\hat{\mbox{\boldmath$v$}} \cdot \mbox{\boldmath$r$})\hat{\mbox{\boldmath$v$}} +(\mbox{\boldmath$r$}-(\hat{\mbox{\boldmath$v$}} \cdot \mbox{\boldmath$r$})\hat{\mbox{\boldmath$v$}} )).
\end{equation}

The test particle four-position transformed from the inertial laboratory frame to the (inertial, for the  non-accelerating) field-source particle rest frame, or source rest frame (srf) is thus
\begin{equation}
{\cal R}^{(\text{srf})}  \equiv ({R}^{(\text{srf})}, \mbox{\boldmath$r$}^{(\text{srf})} ), 
\end{equation}
with
\begin{equation}
{R}^{(\text{srf})}  = \gamma R - \gamma \beta(\hat{\mbox{\boldmath$v$}} \cdot \mbox{\boldmath$r$}),
\end{equation}
and 
\begin{equation}
{\mbox{\boldmath$r$}}^{(\text{srf})}  = -\gamma \beta R \hat{\mbox{\boldmath$v$}} + \gamma(\hat{\mbox{\boldmath$v$}} \cdot \mbox{\boldmath$r$})\hat{\mbox{\boldmath$v$}} +(\mbox{\boldmath$r$}-(\hat{\mbox{\boldmath$v$}} \cdot \mbox{\boldmath$r$})\hat{\mbox{\boldmath$v$}} ),  
\end{equation}
or
\begin{equation}
{\mbox{\boldmath$r$}}^{(\text{srf})}  = \mbox{\boldmath$r$}   - \gamma  R \mbox{\boldmath$\beta$} + (\gamma-1)(\hat{\mbox{\boldmath$v$}} \cdot \mbox{\boldmath$r$})\hat{\mbox{\boldmath$v$}}.  
\end{equation}

The field source particle to test particle separation in the source particle rest frame is thus
\begin{equation}
{\mbox{\boldmath$r$}}^{(\text{srf})}  = \mbox{\boldmath$r$} - (1 - \gamma) (\hat{\mbox{\boldmath$v$}} \cdot \mbox{\boldmath$r$})  \hat{\mbox{\boldmath$v$}} - \gamma \beta R  \hat{\mbox{\boldmath$v$}} \end{equation}

Also, 
\begin{widetext}
\begin{equation}
{R}^{(\text{srf})}  = \left[\left(\mbox{\boldmath$r$} - (1 - \gamma) (\hat{\mbox{\boldmath$v$}} \cdot \mbox{\boldmath$r$})  \hat{\mbox{\boldmath$v$}} - \gamma \beta R  \hat{\mbox{\boldmath$v$}}\right) \cdot\left(\mbox{\boldmath$r$} - (1 - \gamma) (\hat{\mbox{\boldmath$v$}} \cdot \mbox{\boldmath$r$})  \hat{\mbox{\boldmath$v$}} - \gamma \beta R  \hat{\mbox{\boldmath$v$}}\right)  \right]^{1/2},      
\end{equation}
which reduces to
\begin{equation}
{R}^{(\text{srf})}  = R\left[1 - \left[1 - \gamma^2 \right] (\hat{\mbox{\boldmath$v$}} \cdot \mbox{\boldmath$n$})^2  - 2(\hat{\mbox{\boldmath$v$}} \cdot \mbox{\boldmath$n$}) {\gamma}^2 \beta + {\gamma}^2 {\beta}^2   \right]^{1/2},   
\end{equation}
so
\begin{equation}
{\mbox{\boldmath$n$}}^{(\text{srf})}  = \frac{\mbox{\boldmath$r$} - (1 - \gamma) (\hat{\mbox{\boldmath$v$}} \cdot \mbox{\boldmath$r$})  \hat{\mbox{\boldmath$v$}} - \gamma \beta R  \hat{\mbox{\boldmath$v$}}}{ R\left[1 - \left[1 - \gamma^2 \right] (\hat{\mbox{\boldmath$v$}} \cdot \mbox{\boldmath$n$})^2  - 2(\hat{\mbox{\boldmath$v$}} \cdot \mbox{\boldmath$n$}) {\gamma}^2 \beta + {\gamma}^2 {\beta}^2 \right]^{1/2} },      
\end{equation}
or,
\begin{equation}
{\mbox{\boldmath$n$}}^{(\text{srf})}  = \frac{\mbox{\boldmath$n$} - \gamma \mbox{\boldmath$\beta$} - (1 - \gamma) (\hat{\mbox{\boldmath$v$}} \cdot \mbox{\boldmath$n$})  \hat{\mbox{\boldmath$v$}}}{ \left[1 - \left[1 - \gamma^2 \right] (\hat{\mbox{\boldmath$v$}} \cdot \mbox{\boldmath$n$})^2  - 2(\hat{\mbox{\boldmath$v$}} \cdot \mbox{\boldmath$n$}) {\gamma}^2 \beta + {\gamma}^2 {\beta}^2 \right]^{1/2}} .
\end{equation}

Using results above,
\begin{equation}
\frac{\gamma{\mbox{\boldmath$n$}}^{(\text{srf})} }{\left[R^{(\text{srf})}\right]^2} =  \frac{\gamma\mbox{\boldmath$n$} - {\gamma}^2 \mbox{\boldmath$\beta$} - (\gamma - {\gamma}^2) (\hat{\mbox{\boldmath$v$}} \cdot \mbox{\boldmath$n$})  \hat{\mbox{\boldmath$v$}}}{ R^2 \left[1 - \left[1 - \gamma^2 \right] (\hat{\mbox{\boldmath$v$}} \cdot \mbox{\boldmath$n$})^2  - 2(\hat{\mbox{\boldmath$v$}} \cdot \mbox{\boldmath$n$}) {\gamma}^2 \beta + {\gamma}^2 {\beta}^2 \right]^{3/2} }, 
\end{equation}
which can be re-arranged to obtain
\begin{equation}
\frac{\gamma{\mbox{\boldmath$n$}}^{(\text{srf})} }{\left[R^{(\text{srf})}\right]^2} =  \frac{\mbox{\boldmath$n$} - {\gamma} \mbox{\boldmath$\beta$} - (1 - {\gamma}) (\hat{\mbox{\boldmath$v$}} \cdot \mbox{\boldmath$n$})  \hat{\mbox{\boldmath$v$}}}{ {\gamma_s}^2 R^2 \left[1  - (\mbox{\boldmath$\beta$} \cdot \mbox{\boldmath$n$}) \right]^{3}}. 
\label{n_over_Rsqd}
\end{equation}

Eq. (\ref{n_over_Rsqd}) is the needed result for the analysis of Appendix A.  It is perhaps worth noting that it can be further employed to obtain exactly the non-radiative electric field in the lab frame as  
\begin{equation}
\frac{\mbox{\boldmath$E$}}{q} = \frac{\gamma{\mbox{\boldmath$n$}}^{(\text{srf})} }{\left[R^{(\text{srf})}\right]^2} - \frac{\gamma^2}{\gamma  + 1}\frac{\left[\left(\mbox{\boldmath$\beta$} \cdot\mbox{\boldmath$n$}^{(\text{srf})}\right) \mbox{\boldmath$\beta$}\right] }{\left[R^{(\text{srf})}\right]^2} =  \frac{\mbox{\boldmath$n$} - {\gamma} \mbox{\boldmath$\beta$} - (1 - {\gamma}) (\hat{\mbox{\boldmath$v$}} \cdot \mbox{\boldmath$n$})  \hat{\mbox{\boldmath$v$}}}{ {\gamma}^2 R^2 \left[1  - (\mbox{\boldmath$\beta$} \cdot \mbox{\boldmath$n$}) \right]^{3} }   - \frac{\gamma^2}{\gamma  + 1}\frac{\left[\left(\mbox{\boldmath$\beta$} \cdot\mbox{\boldmath$n$}^{(\text{srf})}\right) \mbox{\boldmath$\beta$}\right] }{\left[R^{(\text{srf})}\right]^2},
\end{equation}
\end{widetext}
which under further algebraic manipulations becomes
\begin{equation}
\frac{\mbox{\boldmath$E$}}{q} =  \frac{\mbox{\boldmath$n$} - \mbox{\boldmath$\beta$}}{{\gamma}^2 R^2\left[1  - (\mbox{\boldmath$\beta$} \cdot \mbox{\boldmath$n$}) \right]^{3} }.
\end{equation}


\section{Lorentz Transformation of an Arbitrary Four-Position from an Inertial Frame to the Test Particle Momentary Rest Frame, with a Translation of Origin}

Consider an arbitrarily located particle position \((\mbox{\boldmath$r$}_p(t)\) expressed in an inertial reference frame coordinate system \((x_1,x_2,x_3,t)\), where \(t\) is the time coordinate, with space origin fixed with respect to the laboratory frame observer, and an inertial reference frame coordinate system \((\sigma_1,\sigma_2,\sigma_3,\xi)\) with space origin momentarily co-moving with the test particle at \(t = t_0\), and space origin co-located with the field source particle at \(t = t_0\). The arbitrarily located particle four-position in the lab frame is then 
\begin{equation}
{\cal R}_p = (ct, \mbox{\boldmath$r$}_p(t)= (c t, \mbox{\boldmath$r$}_p).
\end{equation}

Suppose the test particle position as a function of time is \(\mbox{\boldmath$r$}(t)\). The arbitrary four position with space part divided into components parallel and perpendicular to the test particle velocity \(\mbox{\boldmath$v$}(t) = d\mbox{\boldmath$r$}(t)/dt   \) at time \(t=t_0\),  \(\mbox{\boldmath$v$}_{0}\), is then
\begin{equation}
{\cal R}_p  =  (ct ,(\hat{\mbox{\boldmath$v$}}_{0} \cdot \mbox{\boldmath$r$}_p)\hat{\mbox{\boldmath$v$}}_{0} +(\mbox{\boldmath$r$}_p-(\hat{\mbox{\boldmath$v$}}_{0} \cdot \mbox{\boldmath$r$}_p)\hat{\mbox{\boldmath$v$}}_{0})).
\end{equation}

The arbitrary four position transformed to the test particle momentary rest frame (tmrf), relative to the test particle located at the position \(\mbox{\boldmath$r$}(t=t_0) \equiv \mbox{\boldmath$r$}_0\) is  
\begin{equation}
P_p  \equiv {{\cal R}}^{(\text{tmrf})} \equiv (c\xi, \mbox{\boldmath$\sigma$}_p) 
\end{equation}
with
\begin{equation}
\xi  = \xi(t,\mbox{\boldmath$r$}_p(t))  = \gamma_{0} (t - t_0) - \gamma_{0} \beta_{0}(\hat{\mbox{\boldmath$v$}}_{0} \cdot (\mbox{\boldmath$r$}_p(t) - \mbox{\boldmath$r$}_{0}))/c 
\nonumber
\end{equation}
with \(\beta_{0} \equiv v_{0}/c\), and 
\begin{widetext}
\begin{equation}
{\mbox{\boldmath$\sigma$}}_p  = -\gamma_{0} \beta_{0} c (t - t_0) \hat{\mbox{\boldmath$v$}}_{0} + \gamma_{0}(\hat{\mbox{\boldmath$v$}}_{0} \cdot (\mbox{\boldmath$r$}_p(t) - \mbox{\boldmath$r$}_{0}))\hat{\mbox{\boldmath$v$}}_{0} + (\mbox{\boldmath$r$}_p(t) - \mbox{\boldmath$r$}_{0}) - (\hat{\mbox{\boldmath$v$}}_{0} \cdot (\mbox{\boldmath$r$}_p(t) - \mbox{\boldmath$r$}_{0}))\hat{\mbox{\boldmath$v$}}_{0} ),  
\nonumber
\end{equation}
or
\begin{equation}
\mbox{\boldmath$\sigma$}_p  = -\gamma_{0} (t-t_0) \mbox{\boldmath$v$}_{0} + (\gamma_{0}-1)(\hat{\mbox{\boldmath$v$}}_{0} \cdot (\mbox{\boldmath$r$}_p(t) - \mbox{\boldmath$r$}_{0}))\hat{\mbox{\boldmath$v$}}_{0} + (\mbox{\boldmath$r$}_p(t) - \mbox{\boldmath$r$}_{0}),  
\end{equation}
or, with \(\gamma \approx 1 + \beta^2/2 \), 
\begin{equation}
\mbox{\boldmath$\sigma$}_p  \approx -\gamma_{0} (t-t_0) \mbox{\boldmath$v$}_{0} + {\beta_0}^2(\hat{\mbox{\boldmath$v$}}_{0} \cdot (\mbox{\boldmath$r$}_p(t) - \mbox{\boldmath$r$}_{0}))\hat{\mbox{\boldmath$v$}}_{0}/2 + (\mbox{\boldmath$r$}_p(t) - \mbox{\boldmath$r$}_{0}),  
\end{equation}
or,
\begin{equation}
\mbox{\boldmath$\sigma$}_p \approx \mbox{\boldmath$r$}_p(t) - \mbox{\boldmath$r$}_{0} + (\mbox{\boldmath$\beta$}_{0} \cdot (\mbox{\boldmath$r$}_p(t) - \mbox{\boldmath$r$}_{0})) \mbox{\boldmath$\beta$}_{0}/2  - (t-t_0) \mbox{\boldmath$v$}_{0}. 
\end{equation}
\end{widetext}

\bibliographystyle{plain}


\end{document}